\documentclass[journal]{IEEEtran}
\usepackage{amsmath, amsfonts}
\usepackage{nicematrix}
\usepackage{bm}
\usepackage{booktabs}
\usepackage{subfigure}
\ifCLASSINFOpdf
\else
\fi

\begin{document}

\title{Evolutionary Dynamics Based on Reputation in Networked Populations with Game Transitions}

\author{Yuji~Zhang,
        Minyu~Feng,~\IEEEmembership{Senior Member,~IEEE,}
        J{\"u}rgen Kurths,
        Attila Szolnoki
\thanks{Corresponding author: Minyu~Feng (email: myfeng@swu.edu.cn).}
\thanks{Yuji Zhang and Minyu Feng are with the College of Artificial Intelligence, Southwest University, Chongqing 400715, China. }
\thanks{Jürgen Kurths is with the Potsdam Institute for Climate Impact Research, 14437 Potsdam, Germany, and also with the Institute of Physics, Humboldt University of Berlin, 12489 Berlin, Germany.}
\thanks{Attila Szolnoki is with the Institute of Technical Physics and Materials Science, Centre for Energy Research, H-1525 Budapest, Hungary.}
}

\maketitle

\begin{abstract}
The environment undergoes perpetual changes that are influenced by a combination of endogenous and exogenous factors. Consequently, it exerts a substantial influence on an individual's physical and psychological state, directly or indirectly affecting the evolutionary dynamics of a population described by a network, which in turn can also alter the environment. Furthermore, the evolution of strategies, shaped by reputation, can diverge due to variations in multiple factors. To explore the potential consequences of the mentioned situations, this paper studies how game and reputation dynamics alter the evolution of cooperation. Concretely, game transitions are determined by individuals' behaviors and external uncontrollable factors. The cooperation level of its neighbors reflects individuals' reputation, and further, a general fitness function regarding payoff and reputation is provided. Within the context of the donation game, we investigate the relevant outcomes associated with the aforementioned evolutionary process, considering various topologies for distinct interactions. Additionally, a biased mutation is introduced to gain a deeper insight into the strategy evolution. We detect a substantial increase in the cooperation level through intensive simulations, and some important phenomena are observed, e.g., the unilateral increase of the value of prosocial behavior limits promotion in cooperative behavior in square-lattice networks. 
\end{abstract}


\IEEEpeerreviewmaketitle

\section{Introduction}
\IEEEPARstart{T}{he} emergence and maintenance of cooperative behavior constitute pivotal concerns within the research on collective behavior, which widely exists in nature and human societies. Meanwhile, artificial intelligence researchers claim that it is time to prioritize the development of cooperative intelligence, which can promote mutually beneficial joint action \cite{dafoe2021cooperative}. Nevertheless, the persistence of cooperative behaviors in the real world and the underlying mechanisms that sustain them remain an open problem. According to the theory of natural selection proposed by Darwin, its sustenance seems impossible \cite{darwin2004origin}, which contradicts the wide existence of cooperative behavior in nature. 

The combination of the spatial topology of individuals and evolutionary game theory provides a powerful theoretical framework to explore and solve the conundrum mentioned above. The first and most well-known observation was made in 1992 when Nowak and May introduced the prisoner's dilemma game into a networked population arranged on a square-lattice \cite{nowak1992evolutionary}. Subsequently, the emergence of network science, including small-world networks \cite{watts1998collective} and scale-free networks \cite{barabasi1999emergence}, inspired intensive research activity toward more complex modelling. As a result, various networks have been widely applied to address the problems of collective behavior over the past decades, including simple networks \cite{santos2005scale,mao2021effect,qu2022evolutionary}, multilayer networks \cite{boccaletti2014structure,xiong2024coevolution}, temporal networks \cite{li2020evolution,sheng2023evolutionary,pi2025dynamic}, and higher-order networks \cite{alvarez2021evolutionary,majhi2022dynamics}. These generalized network frameworks provide the groundwork for investigations into other network dynamics, e.g., the propagation of epidemics \cite{li2025epidemic} and opinions \cite{banez2022modeling}, making them no longer limited to traditional pairwise and static interactions.

Correspondingly, in recent decades, there has been a surge of theoretical research focusing on the evolutionary origins of cooperation and the dynamics underlying its propagation. In parallel, Nowak has named five key mechanisms to explain the underlying factors behind this enigma \cite{nowak2006five}. Among the mechanisms that have been widely discussed in the context of promoting cooperation, spatial structure not only stands out as a prominent perspective from the field of evolutionary game theory, but also plays a significant role in neutral evolution due to its asymmetry \cite{allen2015molecular}. Not long afterward, Ohtsuki {\it et al.} provided a simple but intuitive rule, i.e., $b/c>k$ (the ratio of the benefit $b$ from cooperation to its corresponding cost $c$ exceeding the average degree $k$), for the threshold of the dominance of cooperative behavior \cite{ohtsuki2006simple}. Notably, Allen and his colleagues provided a general solution applicable to any connected network under weak selection \cite{allen2017evolutionary}. Then, analytical results have been obtained for multilayer networks \cite{su2022evolution}, temporal networks \cite{meng2025promoting}, higher-order networks \cite{sheng2024strategy}, personalized strategy updates \cite{meng2024dynamics}, and self-interaction learning \cite{zeng2025evolutionary}.

It is well-established that both endogenous and exogenous factors and rules contribute to the variability and diversity of interactions and strategies among competitors, thereby exerting a substantial impact on evolutionary dynamics. Coevolutionary rules, as a typical one, have paved a promising avenue for addressing social dilemmas and promoting cooperation \cite{perc2010coevolutionary}. In detail, coevolutionary rules take numerous aspects into consideration, including the aging of individuals \cite{szolnoki2009impact}, teaching activity \cite{szolnoki2008coevolution}, population mobility \cite{xiao2020leaving}, inhibition and activation of interactions \cite{yao2023inhibition}, and other different attributes \cite{li2022graphical} and learning strategy of individuals \cite{zhou2024coevolution,shi2022analysis,zhu2025q}, which allows evolutionary models to describe reality more accurately and effectively. 

The majority of related studies considered single and fixed strengths of dilemmas or were limited to one unchanging game mode in their proposed model. In reality, the situation is often more complicated, as both intrinsic and extrinsic factors continually evolve, altering the environments that individuals encounter over time, which can be validated through human-related activities \cite{rankin2007tragedy} and microbial systems \cite{mcfall2013animals}. Researchers have achieved extensive results regarding how the ever-changing environment affects evolutionary dynamics. Hilbe {\it et al.} introduced the idea that the public resource is dynamic and influenced by the strategic decisions made by individuals \cite{hilbe2018evolution}. They found that cooperation can be remarkably promoted in this setting. Su {\it et al.} conducted comprehensive research on game transitions, including local and global transitions, transitions among $n$ states, and considered the evaluation of the sensitivity of evolutionary dynamics to initial conditions via pair approximation and diffusion approximation \cite{su2019evolutionary}. Feng {\it et al.} researched transitions among different games, including the prisoner's dilemma game, snowdrift game, and stag-hunt game based on a continuous Markov chain, shedding light on how transitions among different game modes affect evolutionary dynamics \cite{feng2023evolutionary}. Different from its positive impact on cooperation, varying environments can also lead to the collapse \cite{stewart2014collapse} and oscillations \cite{weitz2016oscillating} of cooperation. Hence, how changing environments affect evolutionary dynamics has become the most intensively studied research path in evolutionary game theory.

Meanwhile, reputation serves as a pivotal mechanism fostering cooperation and overcoming social dilemmas in human societies \cite{xia2023reputation}, of which typical examples are image scoring \cite{nowak1998evolution}, stern judging, \cite{panchanathan2003tale}, and shunning \cite{tanabe2013indirect}. Significant progress has been made in this
field: in recent related investigations, Zhang {\it et al.} explored the impact of asymmetric fitness comparison based on reputation \cite{zhang2022reputation}. Furthermore, Feng and his colleagues investigated a reputation-driven imitation rule, where the individual's own strategy determines whether its reputation increases \cite{feng2023evolutionary}. One adaptive adjustment of edge weight determined by reputation allows the prosperity of cooperative behavior \cite{li2019reputation}. Hu {\it et al.} proposed a trust game and found that trust flourishes in networks if the reputation threshold is high \cite{hu2021adaptive}. Second-order reputation was considered, and cooperation can be enhanced with the increase of reputation strength \cite{li2022n}. Cuesta {\it et al.} verified the unique role of reputation in fostering cooperation through human experiments \cite{cuesta2015reputation}. In the latest work, adaptive reputation based on the snowdrift game promotes cooperation on simplicial complexes \cite{zhu2025adaptive}. 

Different from previous works, one of the key goals of our present model is to explore how the above-mentioned two mechanisms would affect the outcomes of the evolution and whether defectors could exploit cooperators utilizing their reputations in varying environments under certain situations. Consequently, we here consider the case where the increase and decrease of the individual's reputation is reflected and determined by its neighbors' behaviors, which can be regarded as homophily in social networks \cite{mcpherson2001birds} or neighborhood effects \cite{helms2012keeping,tran2020my}, which encapsulate how human behavioral outcomes is affected by spatial contexts through interconnected social and environmental pathways. That is to say, an individual frequently surrounded by cooperative neighbors enjoys a higher reputation and vice versa. The rules of game transitions are from two aspects: one depends on individuals' behavior, and another is subject to exogenous changes, which indicates that it is independent of the evolutionary dynamics. In previous studies, researchers predominantly presumed a strategy update rule that excluded further microscopic influences. However, the wide existence of microscopic effects, like mutation (also known as random strategy exploration \cite{traulsen2009exploration}) and migration, resulting in uncertainty in strategy selection, has already received substantial attention \cite{santos2016evolution, erovenko2019effect}. Inspired by these observations, we introduce a mutation mechanism for some necessary discussions. Additionally, without losing generality, we suppose that the mutation can exhibit bias, such that the probability of transitioning to cooperation or defection is unequal.

In general, the main contributions of this article are summarized as follows.

1) Based on discrete Markov chains and Geometric Brownian Motion (GBM), we utilize stochastic games to capture the evolving psychology of individuals and public resources in real-world scenarios, where the games individuals engage in change over time. In detail, individuals' behaviors or the drift and diffusion coefficient in GBM determine the rules of game transitions among different states. 

2) Considering the wide existence of mutant behavior in reality, the strategy mutation is introduced for further research and has studied the consequences of biased mutation on the evolution of cooperative behavior.

3) Diverse game transition modes are introduced for comprehensive and in-depth research, encompassing both deterministic and probabilistic game transitions that arise from both endogenous and exogenous factors. 

The structure of the paper is organized as follows. Section~II introduces the strategy evolution dynamics and a reputation-driven mechanism with stochastic properties. In Section~III, we illustrate the methods and exhibit our simulation results. In Section~IV, we conclude our work, give some discussions, and introduce some future potential perspectives. 

\section{Evolutionary Game In Networked Populations Based On Stochastic Games And Reputations}
In this section, we present an innovative evolutionary game model in structured populations that focuses on game transitions and the reputations of individuals. Specifically, in the real world, individuals do not consistently engage in a single game without any modifications. Instead, they possess a dynamic mechanism that evolves over time, which can be linked to their psychology constantly changing. To delineate this phenomenon, we employ the Markovian framework to model the game transitions from both continuous and discrete time, thereby encapsulating the probabilistic nature of state transitions from one game instance to the next.

\subsection{Game Model}
In this paper, we take the evolutionary prisoner's dilemma game (PDG) (also known as the donation game \cite{sigmund2010calculus}) into account as the framework of our proposed model. For that purpose, we provide a concise overview of the original PDG model wherein participants have two optional strategies: cooperation ($C$) and defection ($D$). If two participants achieve mutual cooperation, both derive the same reward ($\mathcal{R}$); however, they obtain the same punishment ($\mathcal{P}$) for mutual defection. Moreover, for two opposite strategies, the unilateral cooperator receives the sucker's payoff ($\mathcal{S}$), while the other adopting defection gets the temptation ($\mathcal{T}$). In addition, $\mathcal{R}$, $\mathcal{P}$, $\mathcal{S}$ and $\mathcal{T}$ satisfy the following conditions in PDG, i.e., $\mathcal{T}>\mathcal{R}>\mathcal{P}>\mathcal{S}$ and $2\mathcal{R} \geq \mathcal{T}+\mathcal{S}$. According to traditional settings, we set $\mathcal{R}=b-c$, $\mathcal{P}=0$, $\mathcal{S}=-c$ and $\mathcal{T}=b$, where $b>c>0$. In PDG, $D$ is always the best regardless of the opponent's strategy, i.e., $(D, D)$ is the Nash equilibrium in a well-mixed population. Furthermore, the interaction pattern is determined by a networked population, in which nodes represent individuals and edges represent interactions. 

\begin{figure*}[htbp]
    \centering
    \subfigure{
    \includegraphics[scale=0.35]{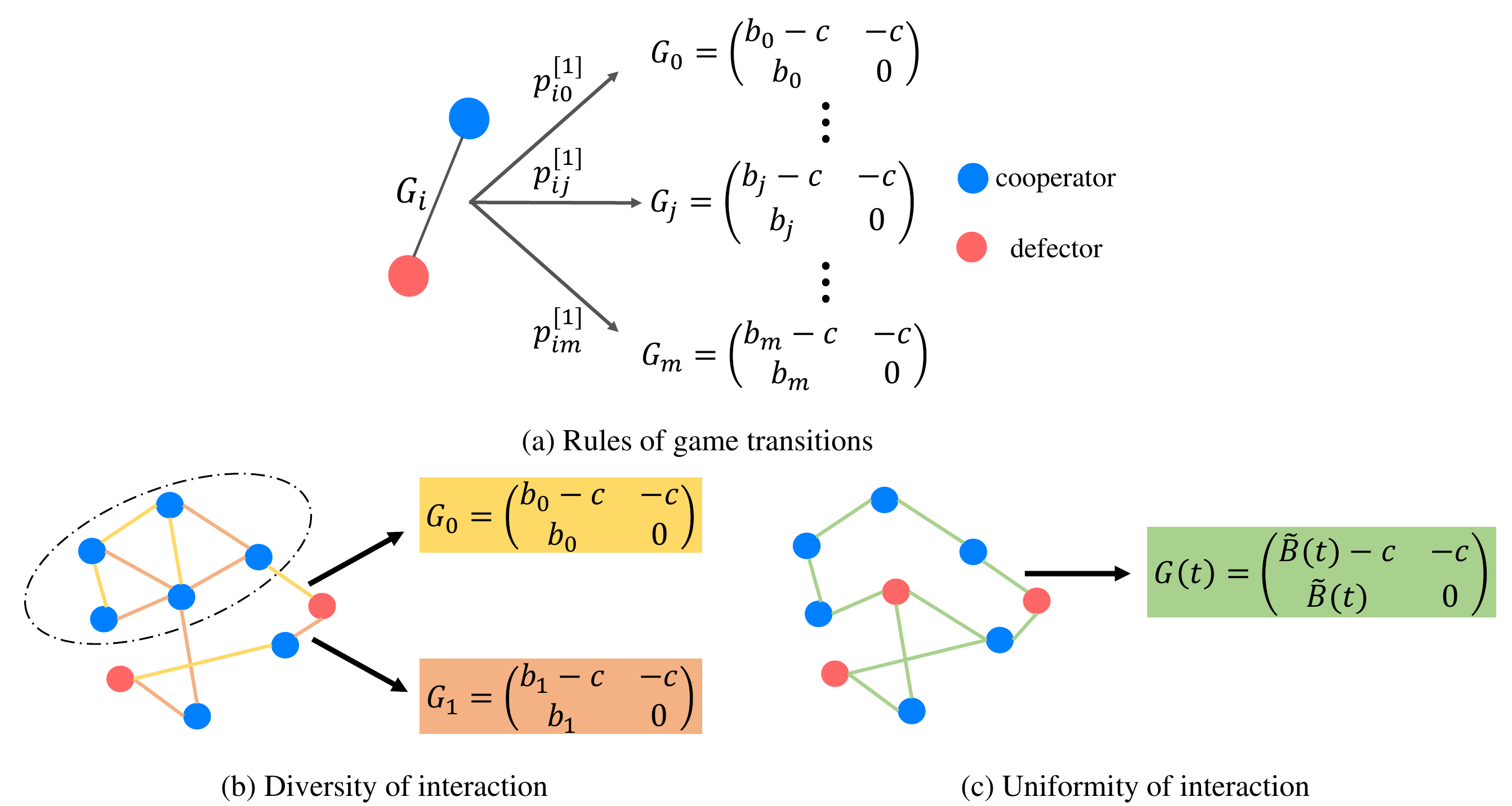}   
    }
    \caption{Illustration of the proposed model. Subplot~(a) exhibits the general rule of the game transition. Concretely, the game state starting from $G_i$, given there is one cooperator, i.e., $l=1$ ($l$ indicating the number of cooperators in each pairwise interaction, the formal definition is given in the main text), has a probability $p^{[1]}_{ij}$ to enter $G_j$, where $\sum_{j \in E}p^{[1]}_{ij} = 1$ for each $G_i \in E$. Subplots (b) and (c) show the game transitions on graphs, where each player has two strategies to choose from, i.e., cooperation (blue) and defection (red), and the edge color indicates the game state. Subplots~(b) and (c) exhibit interactive relationships' diversity and uniformity, respectively. The diversity indicates that each individual can face different social dilemmas at the same time (depicted by different colors in (b)). In contrast, the uniformity means all individuals face the same social dilemma at the same time (depicted by the same color in (c)), and the social dilemma changes as time progresses. Additionally, the dotted circle in subplot~(b) indicates the homophily's effect in the cluster of cooperators. In each time step, players game with their neighbors and accumulate their payoff from all interactions. The diversity in interactions can manifest distinct game types, which are accentuated by the different hues of the connecting edges and their corresponding payoff matrices. At the end of each time step, individuals' strategies update, and then all games and reputations vary according to their corresponding rules. } 
    \label{fig:1}
\end{figure*}

\subsection{Stochastic Games}
To describe the changeable resources in reality, we utilize stochastic games to model them. Hence, we denote the state space $E = \{G_0, \dots, G_m\}$ as all possible $(m+1)$ game states that can be reached during evolution. In the proposed model, we suppose that different game states correspond to different strengths of PDG. Therefore, for PDG and the state $G_i$, the payoff matrix can be expressed as
\begin{equation}
    \begin{pNiceMatrix}[first-row,first-col]
        & C & D \\
    C & b_i-c &   -c \\
    D & b_i & 0 
    \end{pNiceMatrix},
    \label{equation:1}
\end{equation}
in which the benefit parameter $b_i$ can vary during an evolution process. 

Concerning the setting of stochastic games, in a natural way, two different features of game transitions caused by endogenous and exogenous factors are considered. One is dependent on the strategy evolution, and another is independent of the evolutionary process. These include deterministic state-independent and probabilistic behavior-independent game transitions, of which details are listed as follows:

\subsubsection{Deterministic State-Independent Game Transition (DSIGT)}
DSIGT indicates that the game to be played is independent of the previous game and is determined by the games played and the individuals' behaviors taken in the last round. For simplicity, we first take the two game states, $G_0$ and $G_1$, into account. In detail, if there is at least one defector in $G_0$, two individuals will transition to $G_1$ in the next time step. And if they want to enter and remain in $G_0$, they have to adopt and maintain mutual cooperation. If $b_i < b_j$, we say that $G_i$ is ``less valuable'' than $G_j$. Therefore, without loss of generality, we take $b_0 > b_1$ and analyze a natural transition structure where only mutual cooperation brings about the ``most valuable'' game.

The rule of DSIGT can be described by the transition matrix $\mathbf{P}^{[l]}$, where $l = (s_x + s_y) \in \{0,1,2\}$, by mapping $C$ (resp. $D$) to $1$ (resp. $0$), is the number of $C$-players between individuals $x$ and $y$. Concretely, the elements in $\mathbf{P}^{[l]}$ are either $0$ or $1$, i.e., the corresponding transition matrices are
\begin{equation}
\begin{matrix}
\mathbf{P}^{[0]}=\begin{bmatrix}
0  & 1 \\ 
0  & 1 
\end{bmatrix},\,
\mathbf{P}^{[1]} =   
\begin{bmatrix}
0 & 1 \\ 
0 & 1
\end{bmatrix},\,
\mathbf{P}^{[2]} = 
\begin{bmatrix}
1 & 0 \\ 
1 & 0
\end{bmatrix}
\end{matrix}.
\end{equation}

Moreover, in a natural and similar way, we can extend the above case to even more general cases (e.g., probabilistic ones and transitions among more than two states), which can be expressed as  
\begin{equation}
\mathbf{P}^{[l]} = 
\begin{pmatrix}
    p^{[l]}_{00}  & \cdots & p^{[l]}_{0m} \\
    \vdots  & \ddots  & \vdots \\
    p^{[l]}_{m0}  & \cdots & p^{[l]}_{mm} \\
\end{pmatrix}.
\label{general transition matrix}
\end{equation}
Therefore, $p^{[l]}_{ij}$ means the probability of transition from state $G_i$ to $G_j$ in one time step, given there are $l$ $C$-players in the interaction of two individuals. As shown in Fig.~\ref{fig:1}(a), the current game state is $G_i$ and there is one cooperator (the single blue node, $l=1$), and thus the rule of game transition is determined by the elements $p^{[1]}_{ij}$ for $j \in E$ in $\mathbf{P}^{[1]}$. That is, the considered game state will change from $G_i$ to $G_j$ with probability $p^{[1]}_{ij}$. The difference between $G_i$ and $G_j$ lies in the payoff difference for mutual cooperation and unilateral defection (see Eq.~\ref{equation:1}).

In particular, for the case $\mathbf{P}^{[0]}=\mathbf{P}^{[1]}=\mathbf{P}^{[2]}$, the game transition mode can be regarded as exogenous and is not influenced by strategy evolution.

\subsubsection{Probabilistic Behavior-Independent Game Transition (PBIGT)}
PBIGT means that the game to be played depends entirely on the previous game. Furthermore, we consider the continuous time and game state theoretically. Inspired by Geometric Brownian Motion (GBM), we then have the following expression of the revised $\tilde{B}(t)$:
\begin{equation}
   \tilde{B}(t) = B(0)\text{exp}\{(\mu-\frac{1}{2}\sigma^2)t+\sigma W(t)\} + B_1\,,
   \label{GBM}
\end{equation}
where $W(t)$, $\mu$ and $\sigma$ are the standard Brownian Motion, drift coefficient $\mu \in \mathbb{R}$ ($\mathbb{R}$ denotes all real numbers) and scale parameter $\sigma > 0$ respectively. $B(0)$ is the initial value of GBM $B(t)$ at $t=0$ and $B_1$ can be regarded as the lower bound of $\tilde{B}(t)$, i.e., $\tilde{B}(t) \geq B_1$. 

Fig.~\ref{fig:1}(b) exhibits the diversity of game states, i.e., for one individual, its interaction environments with neighbors can be different. As explicitly shown in Fig.~\ref{fig:1}(c), different from Fig.~\ref{fig:1}(b), the game states of all interactions are uniform at the same time step, though game states can change over time. 

GBM has broad applications in Financial Mathematics, e.g., the well-known Black-Scholes Model. Therefore, in the proposed model, it is reasonable for us to utilize the revised GBM $\tilde{B}(t)$ to model the uncontrollable changing dilemma strength over time. 
\subsection{Reputation Mechanism and Strategy Updating Rule}
To better conform to reality, we also take the reputation characteristics $R_i(t)$ of the individuals into account. The reputation of each impacts its fitness directly. Considering the noise and uncertainty during the process, the updating rule of player $i$'s reputation is provided as follows:
\begin{equation}
    R_{i}(t+1) = R_{i}(t) + \xi_{n_c}, \, n_c \in \{0,1,2,\dots,k_i\}\,,
    \label{reputation}
\end{equation}
where $\xi_{n_c}$ follows one probabilistic distribution (e.g., the normal distribution, $\xi_{n_c}\sim N(\mu_{n_c},\sigma^2_{n_c})$ or uniform distribution, $\xi_{n_c}\sim U(a_{n_c}, b_{n_c})$) indicating the randomness during the update, $n_c$ represents the number of cooperators of $i$'s neighbors and $k_i$ denotes the number of $i$'s neighbors (also called the degree of node $i$). In general, it follows the cognition that if one individual has more cooperators around, then it enjoys a higher reputation, and vice versa. This neighbor-dependent reputation mechanism effectively reflects the ``neighborhood effects'', widely existing in societies.

Each player accumulates its payoff by gaming with its neighbors, and its fitness is reflected by its payoff and reputation. Since we do not add any limit in the range of $R_{i}(t)$, we have to implement additional restrictions on the effect of the reputation on fitness. Therefore, we now need one function to map $R_i(t)$ from $\mathbb{R}$ (representing all real numbers) to $(0,a)$, indicating the target function is bounded as an ``S-shaped'' curve. Accordingly, the arctangent function stands as a preferable candidate for achieving this objective, attributable to its advantageous characteristics, such as continuous nature and differentiability, and its wide applications in related works about reputation mechanisms \cite{zhang2021incentive,gkorou2012reducing} as well. After some necessary operations, we arrive at the expression of the fitness $F_i(t)$ of $i$ related to its reputation $R_i(t)$ and payoff $P_i(t)$:
\begin{equation}
    F_{i}(t) = \left\{ \frac{a}{\pi} \text{arctan}\left[ R_{i}(t) \right] + \frac{a}{2} \right\}  \cdot P_{i}(t)\,,
    \label{eq:fitness}
\end{equation}
where $a>0$ and $P_{i}(t)$ is $i$'s accumulated payoff at time $t$ calculated by corresponding payoff matrices following in the conventional setting.

Concerning the imitation process, different from the random choice from neighbors, it is a natural assumption that individuals are more likely to imitate those whose fitness $F$ is higher. Since fitness can be negative, some technical operations are needed. In detail, the imitated player is selected from its neighbors proportional to the adjusted fitness $\overline{F}$, which is calculated by
\begin{equation}
    \overline{F_j} = \frac{F_j - F_{\text{min}}}{F_{\text{max}} - F_{\text{min}}}\,.
\end{equation}
Then, each possible imitated player will be chosen according to their adjusted fitness, where the corresponding probability is proportional to the adjusted fitness. Especially, if all neighbors' fitness is the same, i.e., $F_{\text{max}} = F_{\text{min}}$, then a random one of them will be chosen. The aforementioned rule indicates that: 

(i) The one whose fitness is lowest will not be imitated, indicating the ``survival of the fittest'' to some extent. 

(ii) Higher fitness leads to a higher probability of being imitated.

For the updating rule of strategies of individuals, the same as in most previous studies, we employ the imitation process, where the Fermi function is applied. Concretely, each individual has a greater tendency to imitate its neighbor whose reputation or payoff is higher. Therefore, the probability of individual $x$ learning from the chosen individual $y$ is expressed as follows:
\begin{multline}
        \mathbb{P}(s_{x} \leftarrow s_{y}) = 
   \frac{1-\delta}{1+\text{exp}\{[P_{x}(t) - P_{y}(t)]/\kappa\}} + \\ 
   \frac{\delta}{1+\text{exp}\{[R_{x}(t) - R_{y}(t)]/\kappa\}}\,,
    \label{fermi function}
\end{multline}
where $0 \leq \delta \leq 1$ indicates the strength of reputation and $\kappa$ represents the noise during updating. Concretely, if $\delta = 0$ (resp. $\delta = 1$), the updating process is driven by payoff difference $P_{x}(t) - P_{y}(t)$ (resp. reputation difference $R_{x}(t) - R_{y}(t)$).

\subsection{Rules of Strategy Evolution}
A networked population structure can be denoted by an unweighted and undirected graph $\mathcal{G} = (\mathcal{V},\mathcal{E})$, having $N$ vertices, where $\mathcal{V}$ and $\mathcal{E}$ are the corresponding vertex set and edge set.

Theoretically, in the proposed model, strategy evolution proceeds as a Markov chain, of which the state is the binary vector $\mathbf{s} = (s_1,\cdots,s_N) \in \{C,D\}^{N}$, where $C$ and $D$ denote cooperation and defection, respectively. We assume that the population undergoes discrete-time updates through replacement events. Concretely and without loss of generality, we denote $(\mathfrak{R},\beta)$ as a replacement event, where $\mathfrak{R} \subseteq \{1,\cdots, N\}$ is the set of individuals who tend to update their strategies and $\beta : \mathfrak{R} \rightarrow \{1,\cdots, N\}$ is the mapping between the imitator and the imitated. According to Eq.~\ref{fermi function}, the state of the population at time $t+1$, $\mathbf{s}^{t+1}$ can be described from the current state $\mathbf{s}^{t}$ at time $t$, i.e.,
\begin{equation}
    s_{i}^{t+1}=\begin{cases}
	s_{\beta(i)}^{t}\,,\ i \in \mathfrak{R}\ \mathrm{and}\ \mathrm{with}\ \mathrm{probability}\ \mathbb{P}( s_i\gets s_{\beta(i)}), \\
	s_{i}^{t}\,,\ i \in \mathfrak{R}\ \mathrm{and}\ \mathrm{with}\ \mathrm{probability}\ 1-\mathbb{P}(s_i\gets s_{\beta(i)}),\\
	s_{i}^{t}\,,\ i \notin \mathfrak{R},\\
\end{cases}
\end{equation}
where $s_{\beta(i)}$ is selected from the neighbor set of $i$ according to relevant rules.

Therefore, the strategy evolution is modeled as an evolutionary Markov chain, of which the transition among different states is completely determined by the above rules.

\section{Simulation Results And Discussions}
In this section, we will carry out some computer simulations to exhibit the details and consequences of the proposed model. To elaborate, we initially showcase the methods employed in the subsequent simulations. 
\subsection{Methods}
In this context, we describe the methods used for our subsequent simulations. For the setting of the payoff matrix in Eq.~\ref{equation:1}, as usual, we always set $c=1$ by default and vary $b_i$ to describe the dilemma strength and effect of game transitions. For better presentation of results, we adopt different values of $b_i$ in different simulations, and the actual value is clarified in each simulation. In this work, unless otherwise stated, we let $a=4$ in Eq.~\ref{eq:fitness} and $\kappa=1$ and $\delta = 0.2$ in Eq.~\ref{fermi function}, which account for a suitable relation between $R_i(t)$ and $F_i(t)$, a payoff-reputation weighted strategy updating rule, and a moderate selection strength with a certain noise. To reveal the possible role of the game transition and reputation on cooperative behavior, we utilize different network topologies. In particular, we apply square-lattice networks (SL, having $k=4$ degrees with periodic boundary) and Watts–Strogatz small-world networks (WS, having $k=4$ average degrees, with rewiring probability equal to $0.3$) \cite{watts1998collective}. To facilitate a valid comparison, the entire network always consists of $N=1600$ nodes in each instance, unless otherwise specified. Concerning the parameters related to the reputation mechanism in Eq.~\ref{reputation}, for the details of the variation of reputation $\xi_{n_c}$, we set $\xi_{0}\sim N(-0.2,0.1)$, $\xi_{1}\sim N(0,0.1)$, $\xi_{2}\sim N(0.1,0.1)$, and for $n\geq3$, $\xi_{n}\sim N(0.3,0.1)$ to avoid the excessively rapid and excessively slow growth in reputation, and the initial value of reputation is fixed to $R(0)$, making the coefficient in Eq.~\ref{eq:fitness} equals $1$. In the initial state, cooperators and defectors are evenly distributed throughout the entire population, namely, each individual has an equal probability ($50\%$) to take cooperation or defection, and all game states are fixed to $G_1$ for game transitions between two states. Following the conventional scenario, we assume that in each round all individuals update their strategies synchronously, i.e., $\mathfrak{R} = \{1,\dots, N\}$. The evolutionary steps for all simulations are fixed at $t=2000$, and the final outcomes for each set of parameters are computed by averaging over at least $50$ independent runs to ensure a high degree of accuracy in the simulation results.

\subsection{Evolution of Cooperation Frequency over Time}
\begin{figure}[htbp]
\centering
\subfigure[SL]{
\includegraphics[scale=0.17]{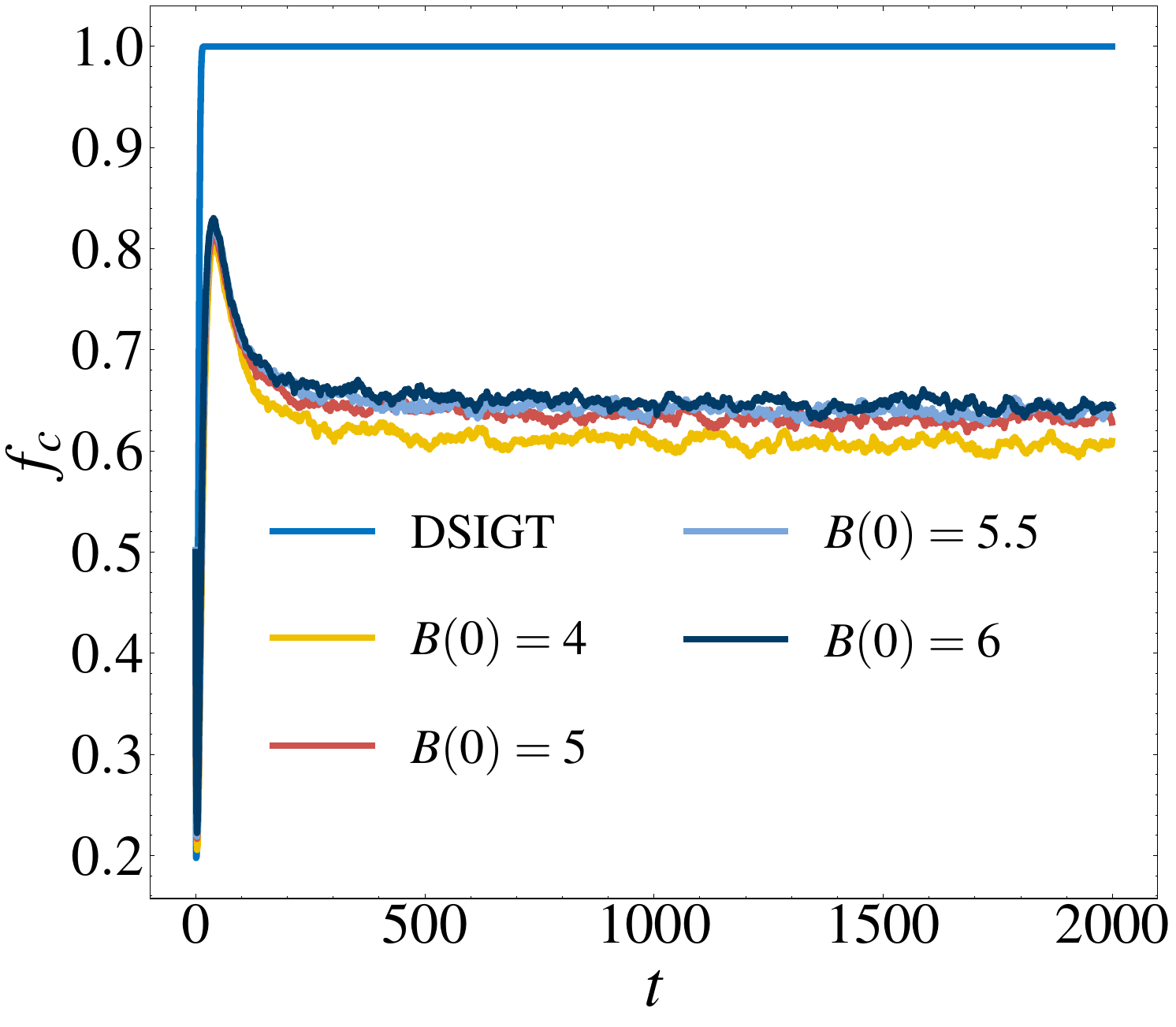}
\label{fig:2 1}
}
\hspace{-5mm}
\subfigure[WS]{
\includegraphics[scale=0.17]{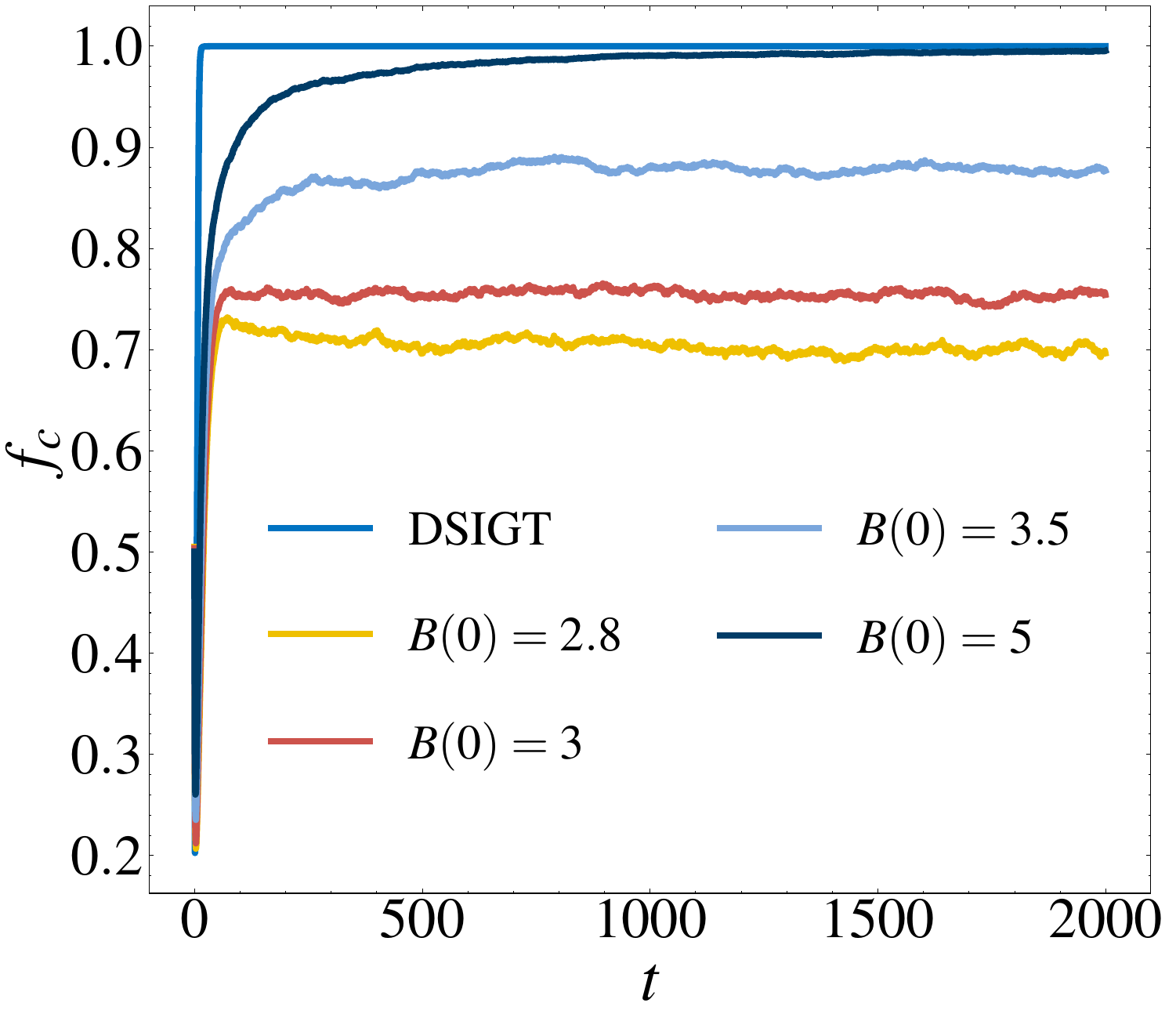}
\label{fig:2 2}
}
\caption{Plots of cooperation frequency against evolutionary time. The ranges of $x$-axis are set as $[0,2000]$ in both subplots, whereas the ranges of $y$-axis in the two subplots are set as $[0.2,1]$. For DSIGTs, it is under the parameter pair $(b_0,b_1) = (6,3)$ in two subplots. Other curves correspond to PBIGTs under $\sigma = 10^{-4}$ and $\mu = \sigma^2 / 2$ with different initial values. Other parameters are: $c=1$, $\delta=0.2$, $a=4$, and $B_1=4$ under SL and $B_1=3$ under WS.}
\label{fig:2}
\end{figure}

For a population, what concerns people most is the cooperation level, which is usually characterized via the cooperation frequency $f_c$, indicating the ratio of cooperators. Therefore, in this subsection, we explore how $f_c$ varies over time $t$ under different conditions, and the relevant results are shown in Fig.~\ref{fig:2}.

As shown in Figs.~\ref{fig:2 1} and \ref{fig:2 2}, for DSIGTs in SL and WS, $f_c$ evolves to $1$ quickly after a sharp decrease initially, indicating its effectiveness in the promotion of cooperative behaviors. However, that is not the case for PBIGTs. Concretely, in SL, by varying the initial value $B(0)$, $f_c$ does not exhibit obvious changes, compared to $f_c$ in WS. In other words, evolution outcomes in SL are insensitive to the initial values of GBM. The insensitivity observed in SL is partly caused by its regular topological structure, which eliminates the impact of the initial values of GBM. Due to the GBM exhibiting a high degree of randomness could result in significant fluctuations in the evolution outcomes. Therefore, we make a preliminary discussion on PBIGT, and in the following subsections, we focus on the endogenous game transitions.

\begin{figure}[htbp]
\centering
\subfigure{
\includegraphics[scale=0.165]{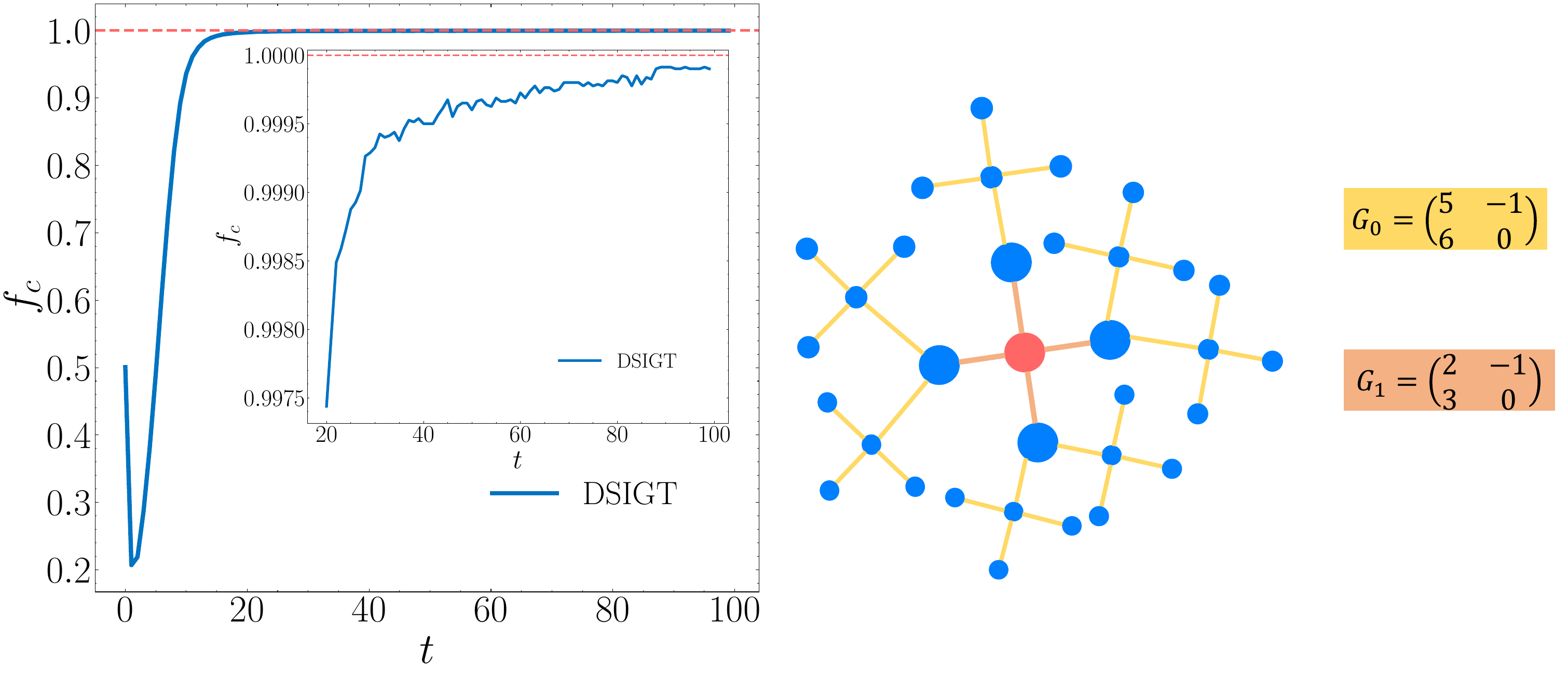}   
}
\caption{Illustration of the particular dilemma. The center player is the defector and the others cooperate. This condition is expected to persist for an extended period, especially under strong selection. Parameters are the same as DSIGTs in Fig.~\ref{fig:2} during simulations.}
\label{fig:3}
\end{figure}

Interestingly, if we change the update rule, i.e., Eq.~\ref{fermi function} from the difference between two individuals' payoff, $P_x(t) - P_y(t)$, to the difference between two individuals' fitness, $F_x(t) - F_y(t)$, after a period of evolution, we observe that in WS, there exists a tiny minority of individuals who adopt unilateral defection for a long time. They inhibit the whole population from evolving into pure cooperators, by taking advantage of their high reputation and payoff (directly leading to higher fitness), which leads to another dilemma, as shown in Fig.~\ref{fig:3}. 

In Fig.~\ref{fig:3}, the focal individual is a defector and the others are cooperators. The defector has four neighbors and according to the rule of DSIGT, the game played by it and its neighbors is $G_1$, which is less valuable than $G_0$. Therefore, the payoff of the focal player is $12$, and the second-nearest neighbors accumulate $24$ units of payoff. According to the rule of imitation, it results in a dilemma, where the nearest neighbors will never imitate the focal player's strategy. In other words, it inhibits the defector from invading the networked population. On the other hand, the defector also can not be occupied by surrounding cooperators, of which the probability, calculated by Eq.~\ref{fermi function},  is extremely low (especially under strong selection, i.e., low noise). However, the population will end up with pure cooperators after a sufficiently long time for evolution. 

In conclusion, it is an interesting phenomenon, indicating the existence of some particular scenarios where individual defectors persist throughout a long period in the evolutionary process. In most cases, though, the timescale of fixation (absorbing) is not concerned, compared to the fixation itself.

\subsection{Difference between Two Games}
\begin{figure}[htbp]
\centering
\subfigure[SL]{
\includegraphics[scale=0.17]{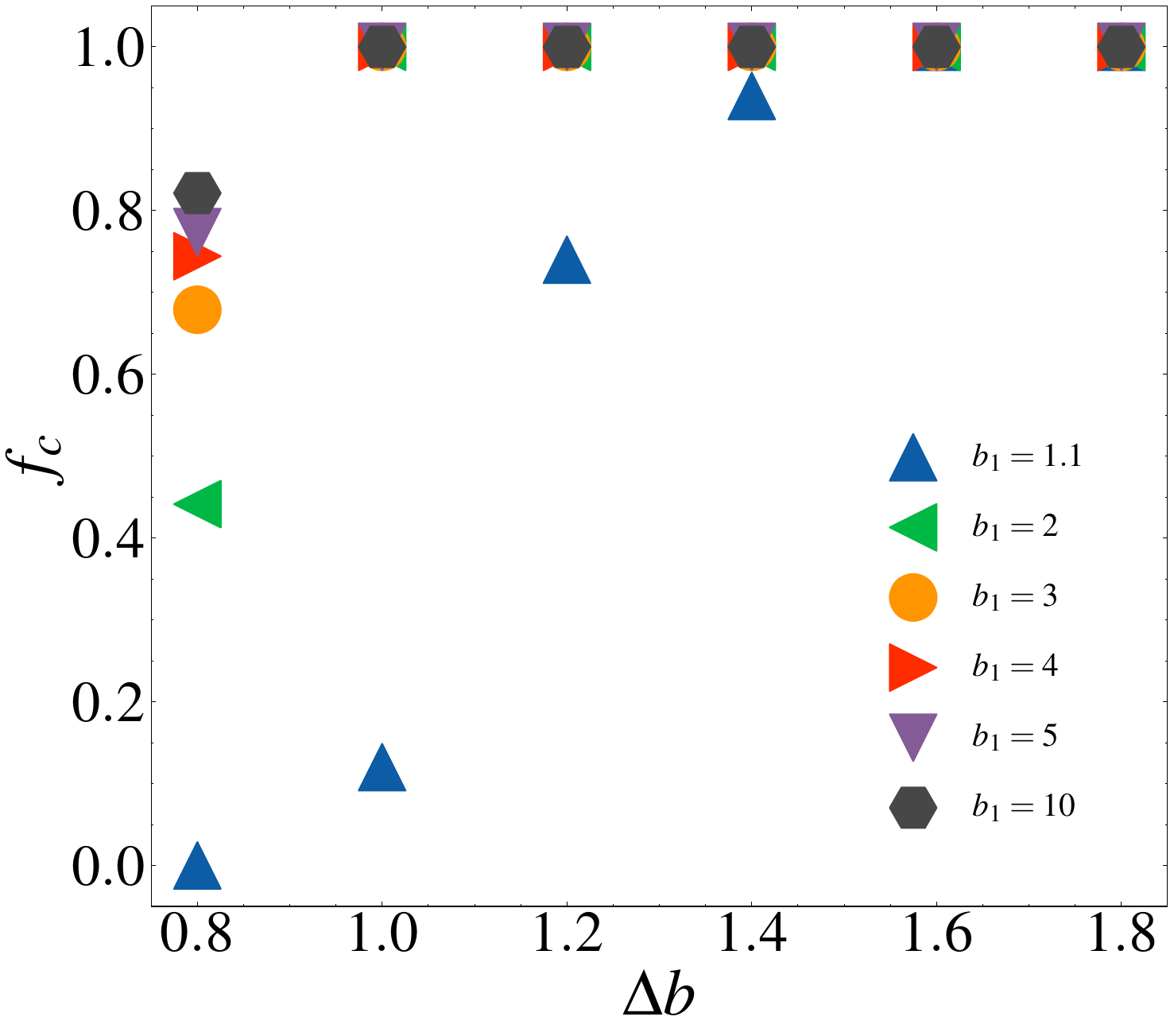}
\label{fig:4 1}
}
\hspace{-5mm}
\subfigure[WS]{
\includegraphics[scale=0.17]{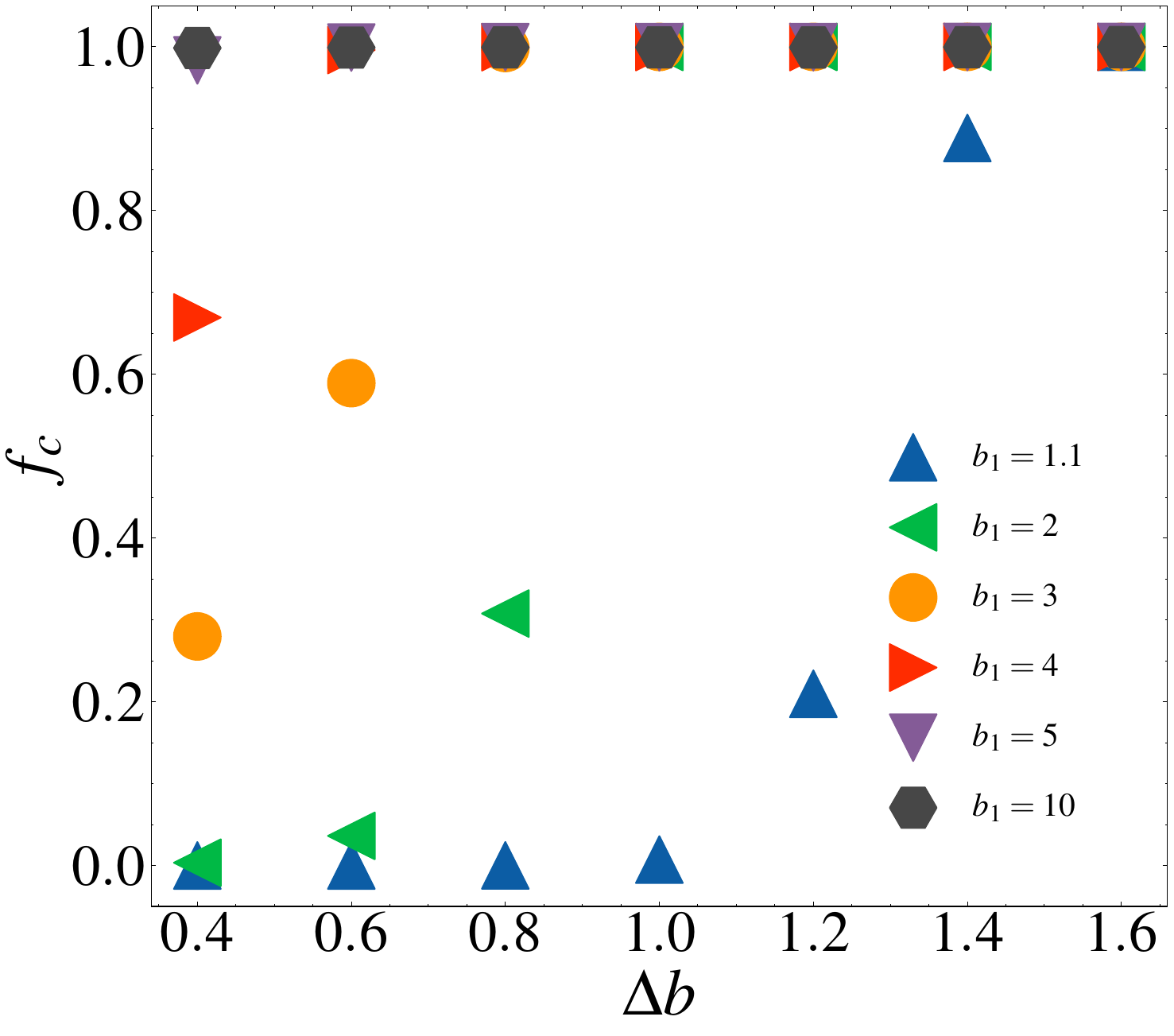} 
\label{fig:4 2}
}
\caption{Plots of cooperation frequency against differences between two games. We set $\Delta b = b_0 - b_1$ and values of $b_1$ in $\{1.1, 2, 3, 4, 5, 10\}$ in SL and WS. The ranges of $x$-axis are set as $[0.8,1.8]$ and $[0.4,1.6]$ in subplots (a) and (b), respectively, with steps equal to $0.2$, whereas the ranges of $y$-axis in the two subplots are set as $[0,1]$. Other parameters are: $c=1$, $\delta=0.2$, $\kappa=1$, and $a=4$.}
\label{fig:4}
\end{figure}

In Ref.~\cite{su2019evolutionary}, under birth-death and pairwise-comparison updating, the condition for $\rho_C > \rho_D$ is $\Delta b>2c$ ($\rho_C$ and $\rho_D$ refer to the fixation probability of cooperation and defection respectively), indicating that the success of cooperation only depends on the difference between the two games (see Eq.~2 and Figs.~3(B) and 3(D) in \cite{su2019evolutionary} for more details). Inspired by this, in the proposed model, we investigate how the difference between two games will affect $f_c$ in SL and WS. Representative results of the proposed model here are demonstrated in Fig.~\ref{fig:4}.

As shown in Figs.~\ref{fig:4 1} and \ref{fig:4 2}, we uncover the rule that in SL and WS, the whole population evolves into the pure cooperators for $\Delta b > 1.6$ under the condition $c=1$. During the simulations, we find that in some cases, the population would quickly evolve into a pure state. For instance, for $b_0 = 2.5$ and $b_1 = 1.1$ ($\Delta b = 1.4$) in SL, the population evolves into pure defectors with a certain probability (about $6\%$). Tab.~\ref{tab:1} gives details of the fixation frequency of cooperation and defection with the increase of simulation times. From this, we can infer that with the increase of $\Delta b$, the fixation times of cooperation increase. In other words, it is more beneficial to the success of cooperation. During simulations, when $\Delta b$ exceeds a certain value, the whole population will quickly evolve into one monomorphic strategic state. Moreover, WS is more sensitive or SL is more robust to the variation of $b_1$, where in SL for $b_1 \in \{3,4,5\}$, the results of $f_c$ exhibit almost no apparent distinction among them. The growth rate of $f_c$ is higher for smaller $b_1$, especially in SL and WS for $\Delta b$ from $1.0$ to $1.2$ and $1.2$ to $1.4$ under $b_1 = 1.1$, respectively. 
\begin{table}[htbp]
  \centering
  \caption{Results of the Frequency of the Whole Population Evolves into Pure States} 
  \begin{tabular}{ccccc} 
    \toprule 
    $b_0$ &  $b_1$ & simulations &  pure cooperators & pure defectors \\
    \midrule 
      $2.5$ &  $1.1$ & $1000$ &  $938$ & $62$ \\
      $2.6$ &  $1.1$ & $1000$ &  $987$ & $13$ \\ 
      $2.7$ &  $1.1$ & $1000$ &  $996$ & $4$ \\ 
      $2.8$ &  $1.1$ & $1000$ &  $998$ & $2$ \\ 
    \bottomrule
  \end{tabular}
  \label{tab:1} 
\end{table}
Overall, we obtain one interesting outcome in SL under reputation and game dynamics by simulations, where the unilateral increase of $b_1$ has a limited effect on promoting cooperation. Moreover, the success of cooperation depends on the $\Delta b$, instead of the exact value in two donation games. The results indicate that game transition can impose a powerful and positive influence on the evolution of cooperation. 

\subsection{Evolution under Different Noise levels}
\begin{figure}[htbp]
\centering
\subfigure[SL]{
\includegraphics[scale=0.17]{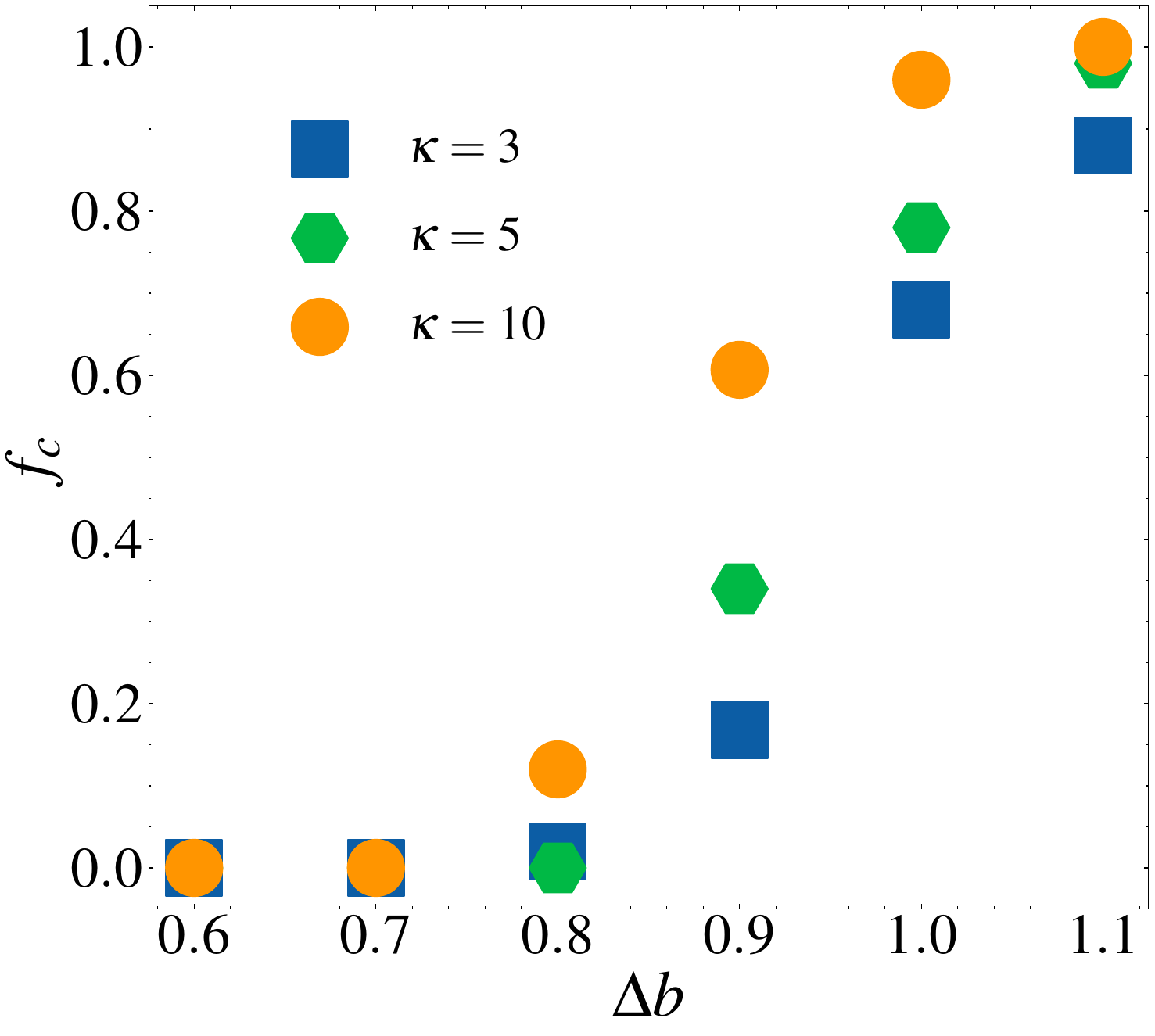}  
\label{fig:5 1}
}
\hspace{-5mm}
\subfigure[WS]{
\includegraphics[scale=0.17]{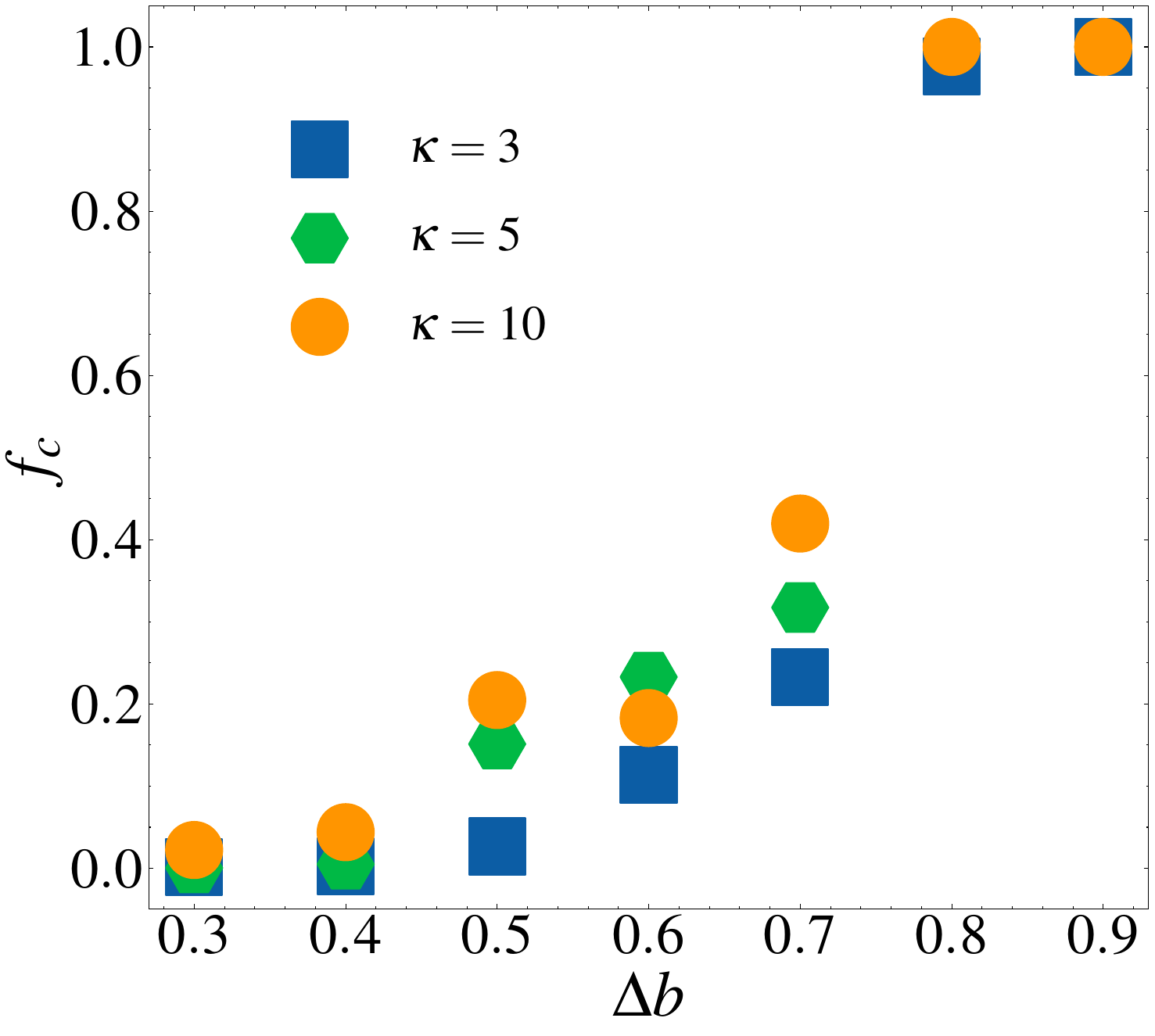}    
\label{fig:5 2}
}
\caption{Plots of cooperation frequency under different noise values $\kappa$. The ranges of $x$-axis are set as $[0.6,1.1]$ and $[0.3,0.9]$ in subplots~(a) and (b), with steps equal to $0.1$, whereas the ranges of $y$-axis in two subplots are set as $[0.0,1.0]$. $b_1$ is fixed as $2$ in both subplots. Other parameters are: $c=1$, $\delta=0.2$, and $a=4$.}
\label{fig:5}
\end{figure}

In this section, we remove the restrictions on the noise parameter $\kappa = 1$ to explore how the evolution of cooperation is affected by noise strengths. In the following numerical analysis, we set three values of $\kappa$, i.e., $\kappa = 3$, $\kappa = 5$, $\kappa = 10$, and corresponding results are exhibited in Fig.~\ref{fig:5}.

According to Fig.~\ref{fig:5} (SL in Fig.~\ref{fig:5 1} and WS in Fig.~\ref{fig:5 2}), we can observe that under larger noise $\kappa$, it is more beneficial for the evolution of cooperation, especially in SL (circles in Fig.~\ref{fig:5 1}). This finding indicates that $\kappa$ plays a positive role in promoting cooperation. It also verifies the robustness of our model.

\subsection{Sensitivity to Initial Condition of Game State}
\begin{figure}[htbp]
\centering
\subfigure[SL]{
\includegraphics[scale=0.17]{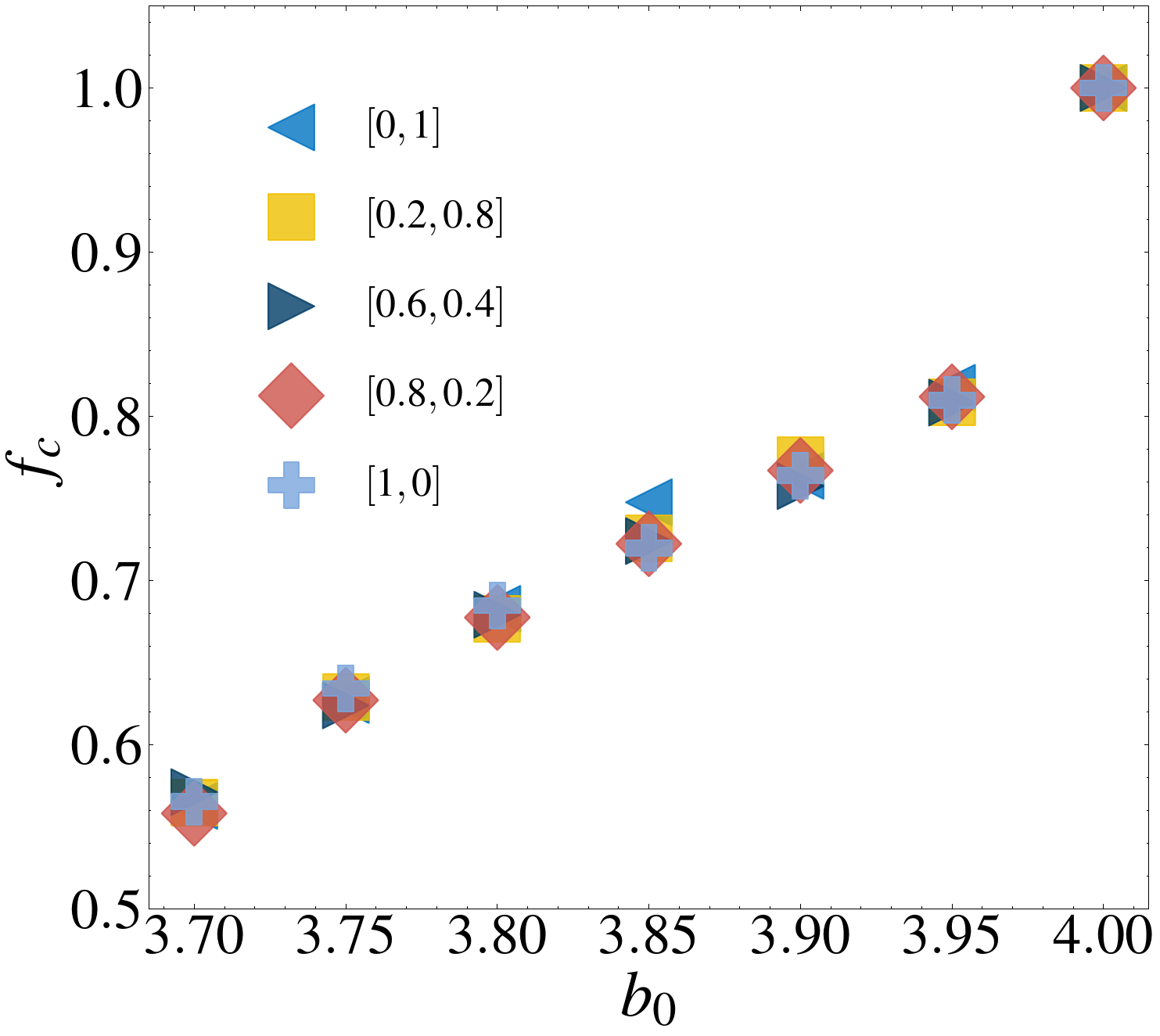}
\label{fig:6 1}
}
\hspace{-5mm}
\subfigure[WS]{
\includegraphics[scale=0.17]{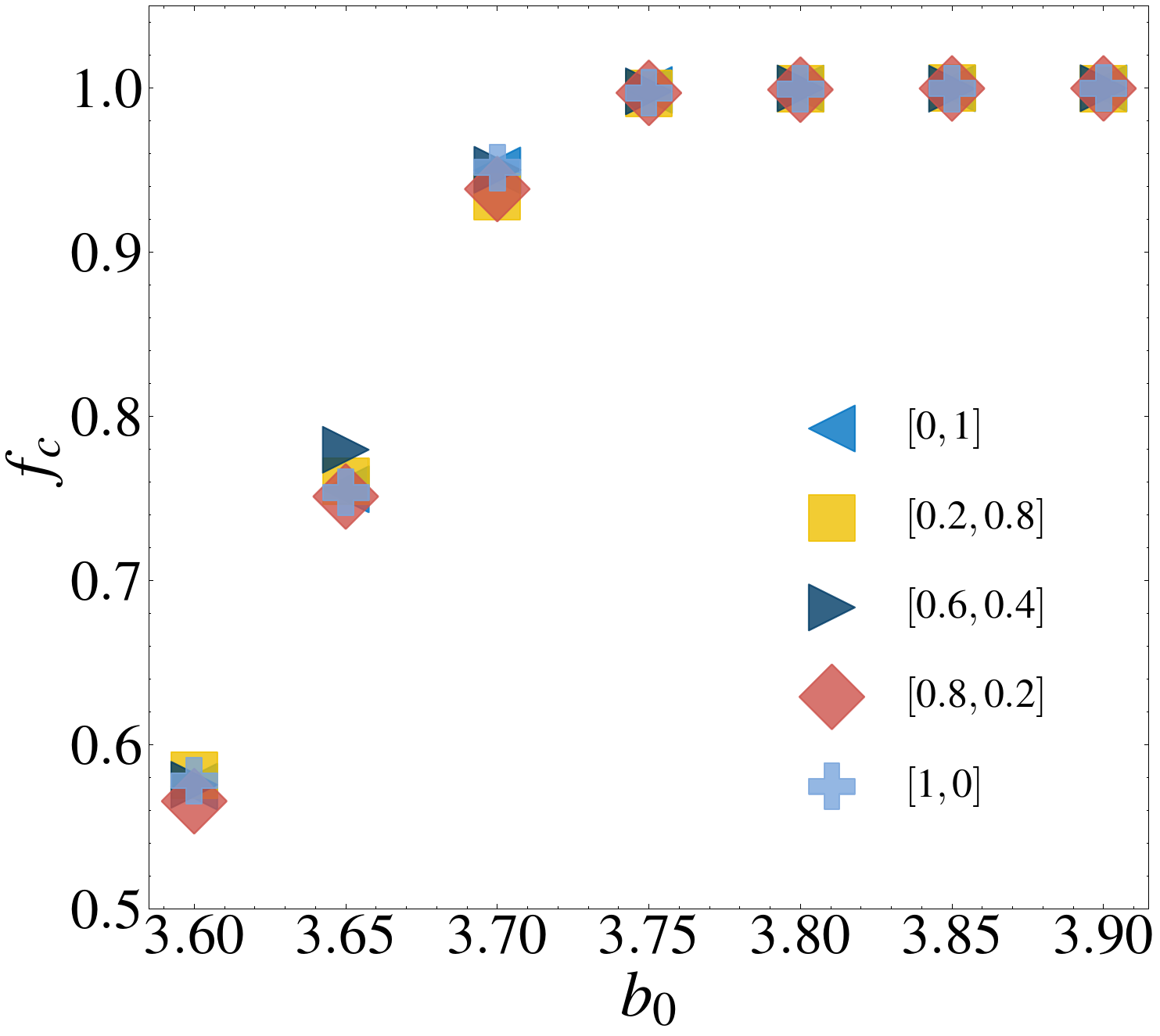}
\label{fig:6 2}
}
\caption{Plots of cooperation frequency against different initial game states. The ranges of $x$-axis are set as $[3.7,4.0]$ and $[3.6,3.9]$ in subplots~(a) and (b), with steps equal to $0.05$. $b_1$ is fixed as $3$ in both subplots. Other parameters: $c=1$, $\delta=0.2$, $\kappa=1$, and $a=4$.}
\label{fig:6}
\end{figure}

In previous computer simulations, we always assumed that each evolution starts from all states equal to $G_1$. It is a natural question to check whether the evolutionary outcome is sensitive to the initial distribution of the game state. Therefore, in this subsection, we explore whether and how the initial condition affects $f_c$, and relevant results are shown in Fig.~\ref{fig:6}. 

In Figs.~\ref{fig:6 1} and \ref{fig:6 2}, we examine different initial patterns, e.g., $[0.2,0.8]$ means during initialization, there is a $20\%$ probability of initializing it as $G_0$ and $80\%$ as $G_1$. As we can clearly see, $f_c$ does not exhibit significant differences under different initial conditions. Moreover, in SL, there is a linear relationship, to some extent, between $f_c$ and $b_0$, for $b_0$ varying from $3.7$ to $3.95$.

Unfortunately, it is almost impossible to conclude that the evolutionary outcome is insensitive to the initial condition from either computational simulations or a theoretical perspective in the proposed model, due to the existence of the unique case below. Now we suppose that $\mathbf{P}^{(s)}$ has the following form
\begin{equation}
\begin{matrix}
\mathbf{P}^{[0]}=\begin{bmatrix}
1  & 0 \\ 
0  & 1 
\end{bmatrix},\,
\mathbf{P}^{[1]} =   
\begin{bmatrix}
1 & 0 \\ 
0 & 1
\end{bmatrix},\,
\mathbf{P}^{[2]} = 
\begin{bmatrix}
1 & 0 \\ 
0 & 1
\end{bmatrix}
\end{matrix}.
\label{special transitions}
\end{equation}
It indicates that the game pattern is fixed and remains the same as the initial state, degenerating into the conventional game mode. Therefore, the outcome of evolution depends on the initial fraction of $G_0$ and $G_1$, where $G_0$ is more valuable than $G_1$ under the previous assumption. The above description naturally brings to mind the mechanism, related to ``social ties'', named ``edge diversity'' (each type of edge representing a specific relationship between two connected individuals) \cite{su2019edgediversity} and connects it to the present case described by Eq.~\ref{special transitions}. 

It is worth noting that for evolutionary dynamics with pure game transitions, the answer is clear \cite{su2019evolutionary}. That is, for game transition matrices in Eq.~\ref{general transition matrix}, the evolutionary outcome is insensitive to the initial condition, but that is not the case for others, e.g., that in Eq.~\ref{special transitions}. Furthermore, in the supporting material of Ref.~\cite{su2019evolutionary}, the authors have provided one effective approach to predict the sensitivity of results.

\subsection{Reputation Distribution of Whole Networked Population}
\begin{figure}[htbp]
\centering
\subfigure[SL]{
\includegraphics[scale=0.165]{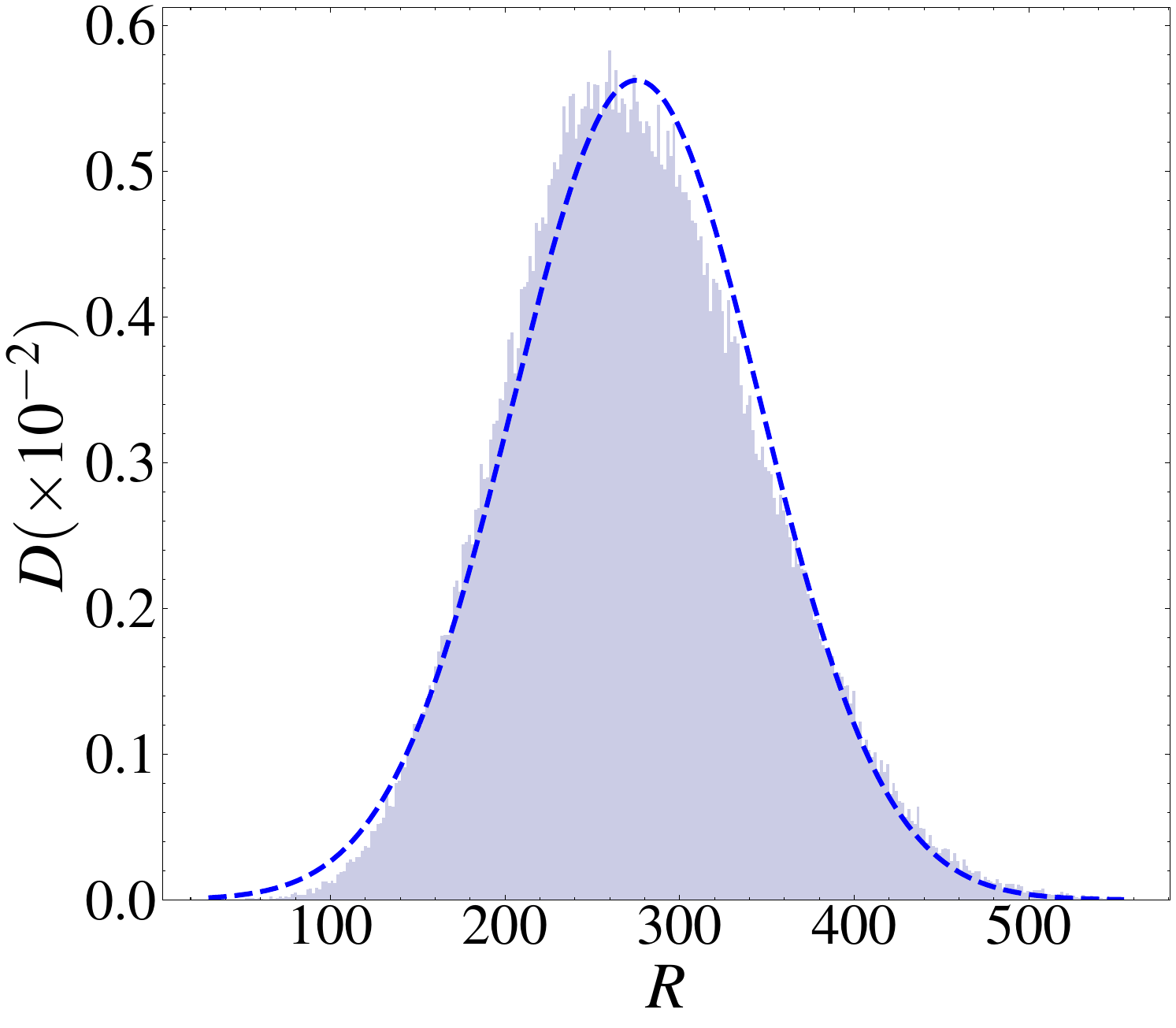}  
\label{fig:7 1}
}
\hspace{-5mm}
\subfigure[WS]{
\includegraphics[scale=0.165]{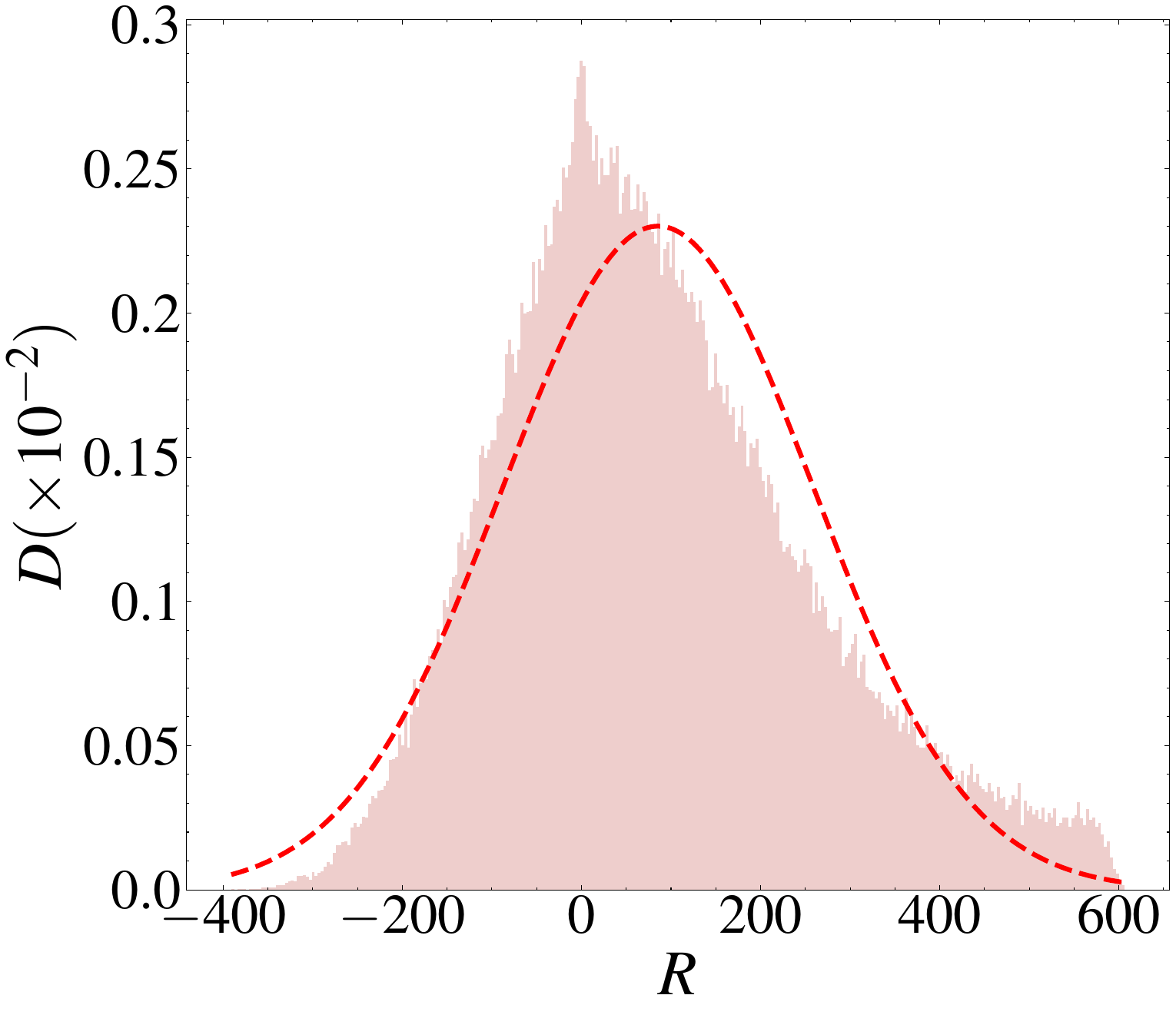}    
\label{fig:7 2}
}
\caption{Distribution of reputation of the system at fixed times. We set $t$ to $2000$ to record and observe the distribution of all individuals' reputations in SL and WS, shown in subplots (a) and (b), respectively. The parameters of the two subplots are set as $b_0=3.7$ and $3.5$ respectively, with the same $b_1 = 3$. The final outcomes are averaged over $100$ independent simulations to maintain a good accuracy of the simulation results. Other parameters are: $c=1$, $\delta=0.2$, $\kappa=1$, and $a=4$.}
\label{fig:7}
\end{figure}

In the proposed model, we display that the reputation accrues over time. Therefore, after some time of evolution, the statistical properties (e.g., the mean and variance) and distribution of the system's reputation intrigue us. The resulting reputation distributions, obtained after $2000$ steps, are shown in Fig.~\ref{fig:7} for different network topologies.

As shown in Fig.~\ref{fig:7 1}, the distribution of the reputation approximates the normal distribution at $t = 2000$ in SL. The mean and standard deviation of the reputation in Figs.~\ref{fig:7 1} and \ref{fig:7 2} are around $(275, 71)$ and $(85, 173)$, respectively. That is to say, compared to SL, the expectation of reputation in WS is smaller, but the variance is higher. In detail, as we can observe in WS, compared to SL, a certain proportion of individuals with reputations lower than zero exists. Interestingly, the observed proportion of such individuals is significantly higher than what is predicted by the normal distribution model fitted to the data, compared to that in SL. Moreover, the overall reputation distribution relatively deviates from the normal distribution in WS, exhibiting characteristics of skewness.

\subsection{Cooperation Frequency on the $(\delta,a)$ Parameter Space}
\begin{figure}[htbp]
\centering
\subfigure[SL]{
\includegraphics[scale=0.17]{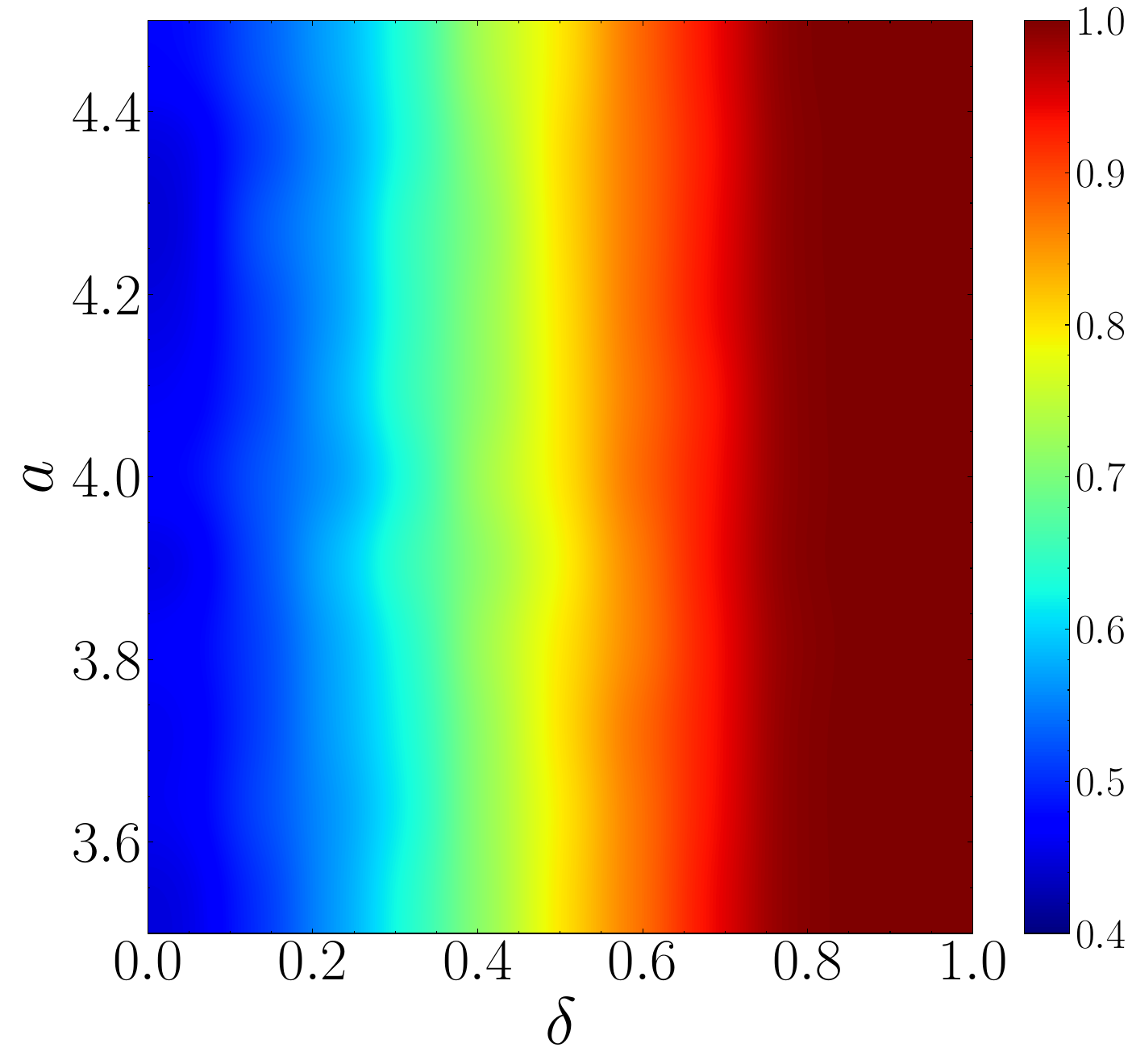}  
\label{fig:8 1}
}
\hspace{-5mm}
\subfigure[WS]{
\includegraphics[scale=0.17]{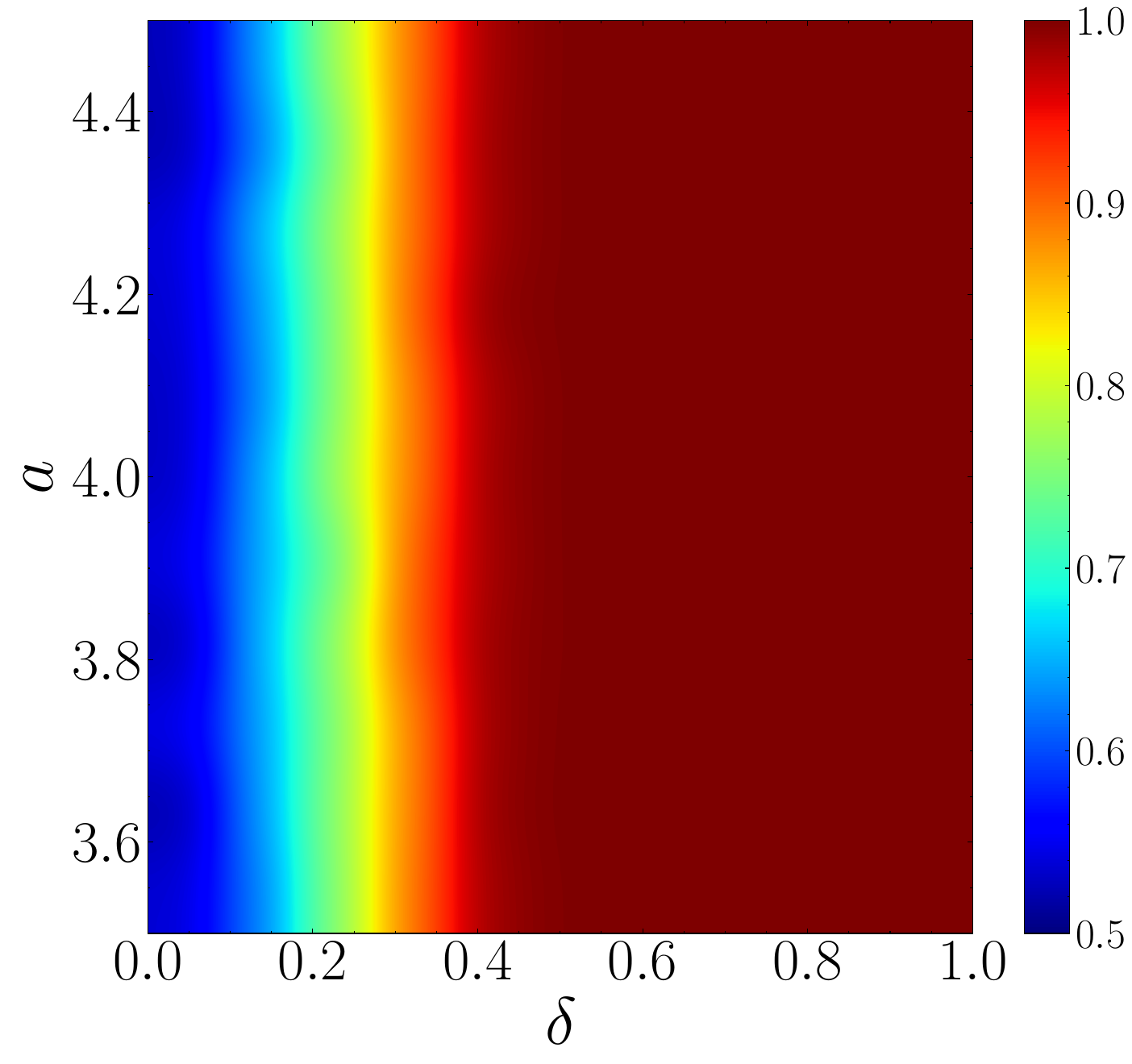}    
\label{fig:8 2}
}
\caption{Heatmaps of cooperation frequency ($f_c$) on $\delta$-$a$ parameter space in SL (panel~(a)) and WS (panel~(b)) networks. The $x$-axis is set as $\delta$ within the range $[0.0,1.0]$ and $y$-axis is set as $a$ within the range $[3.5,4.5]$. The meaning of colors is explained in the right-hand side legend for each panel. The outcomes are averaged over $50$ independent simulations to maintain a good accuracy of the simulation results. Other parameters are: $b_0 = 3.7$ under SL and $b_0=3.65$ under WS, $b_1 = 3$, $c=1$, and $\kappa=1$.}
\label{fig:8}
\end{figure}

In the previous subsections, two important parameters regarding the effect and strength of reputation, i.e., $a$ and $\delta$ are fixed as $4$ and $0.2$, respectively. It remains unclear whether and how $a$ and $\delta$ affect $f_c$ level. Therefore, we made systematic measures at different pairs of these parameters and the results are summarized in Fig.~\ref{fig:8}.

Across the two considered network types (SL in subplot~(a) and WS in subplot~(b)), $f_c$ is observed to increase monotonically with the rise in parameter $\delta$. However, the variation in $a$ does not exert a sensible influence on $f_c$. Generally, in SL and WS, the effect of $a$ and $\delta$ is relatively limited in the promotion of cooperative behavior compared to the difference between the two games $\Delta b$, as shown in previous subsections. And, the variation of $f_c$, under the same parameters, is sharper in WS than in SL, reflected by the range of the color bar. Furthermore, both heatmaps exhibit distinct vertical strip-like patterns, in which the color variations are primarily confined to changes along the $\delta$. This highlights the important role of $\delta$ and reputation in evolutionary outcomes.

\subsection{Game Transitions in General Case}
\begin{figure}[htbp]
\centering
\subfigure[SL]{
\includegraphics[scale=0.17]{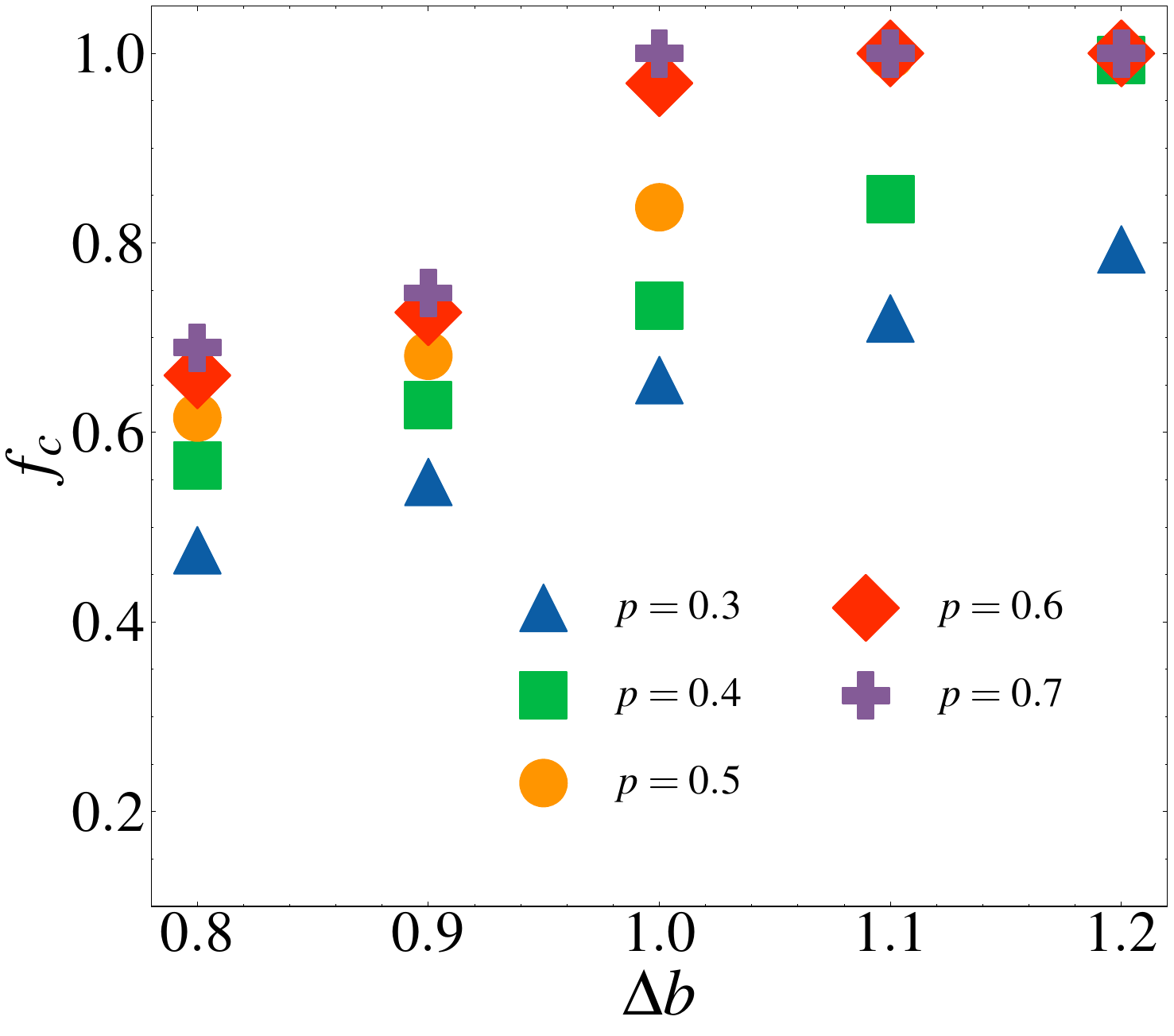}  
\label{fig:9 1}
}
\hspace{-5mm}
\subfigure[WS]{
\includegraphics[scale=0.17]{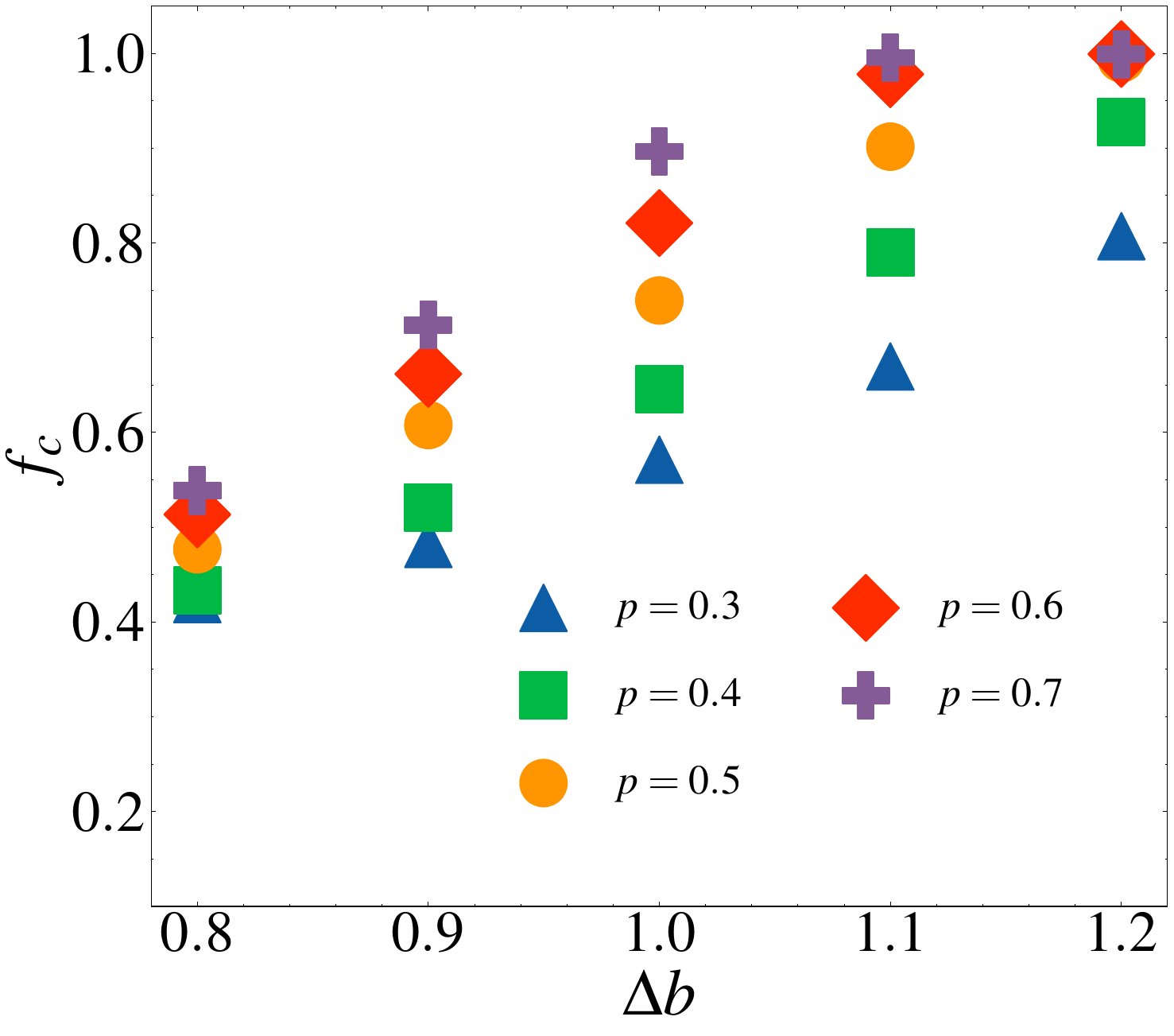}    
\label{fig:9 2}
}
\caption{Plots of cooperation frequency $f_c$ ($y$-axis) against differences between three games $\Delta b$ ($x$-axis). For simplicity, we set $\Delta b = b_0 - b_1 = b_1 - b_2$ and values of $p$ in $\{0.3, 0.4, 0.5, 0.6, 0.7, 0.8\}$ in SL and WS. In two subplots, we fix $b_1$ to $4$. The ranges of $x$-axis are set as $[0.8,1.2]$ with steps equal to $0.1$ in two subplots, with steps equal to $0.2$. Other parameters are: $c=1$, $\delta=0.2$, $\kappa=1$, and $a=4$.}
\label{fig:9}
\end{figure}

\begin{figure}[htbp] 
\centering
\subfigure[SL]{
\includegraphics[scale=0.165]{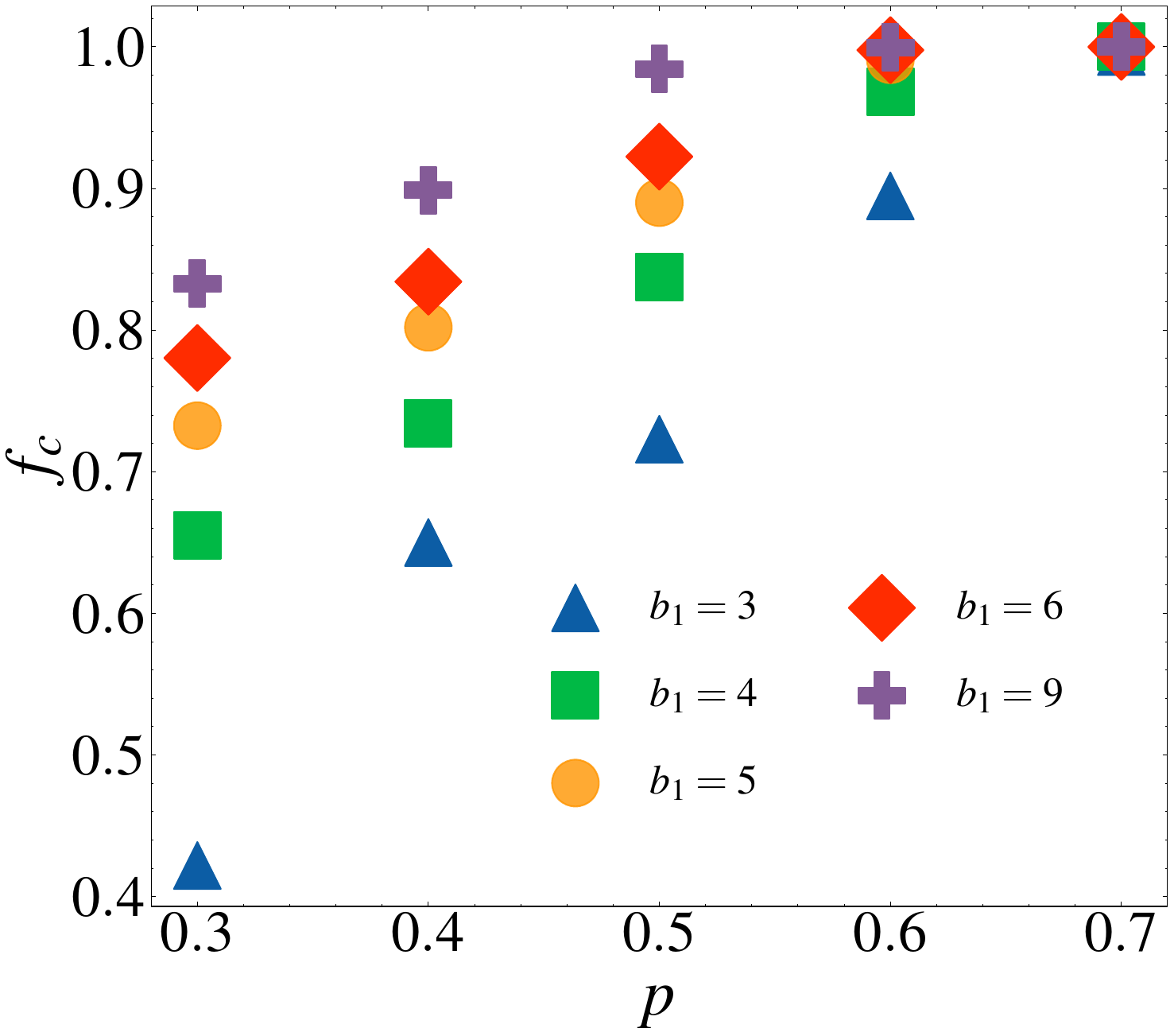}  
\label{fig:10 1}
}
\hspace{-5mm}
\subfigure[WS]{
\includegraphics[scale=0.165]{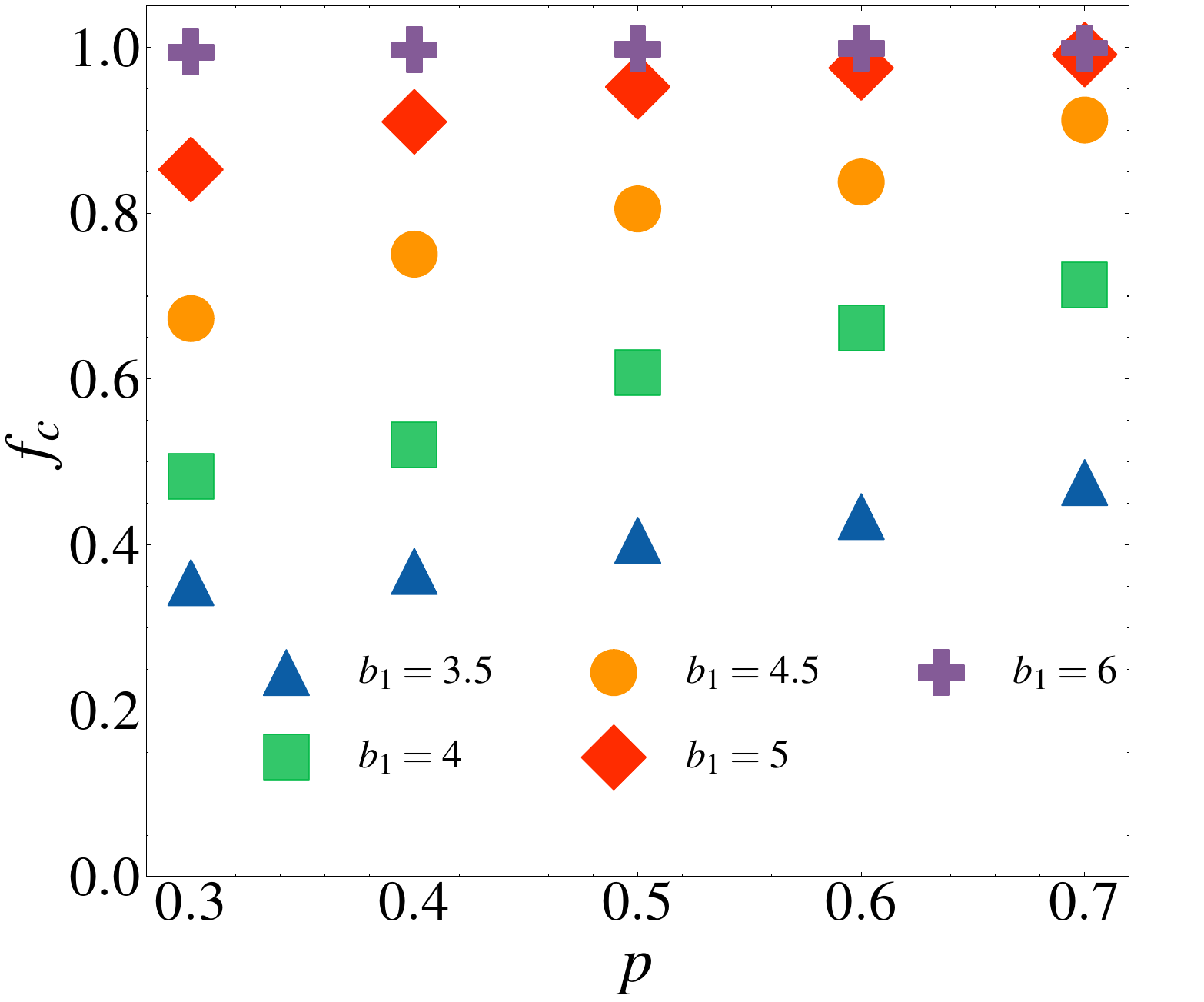}
\label{fig:10 2}
}
\caption{Plots of cooperation frequency $f_c$ ($y$-axis) against different transition probabilities $p$ ($x$-axis). For simplicity, we set $\Delta b = b_0 - b_1 = b_1 - b_2 $ equal to $1$ and $0.5$ in SL and WS respectively (e.g., the legend $b_1 = 3$, in subplot~(a), indicates $b_0 = 4$ and $b_2 = 2$ and in subplot~(b), the legend $b_1 = 3.5$ means $b_0 = 4$ and $b_2 = 3$) and make values of $p$ vary from $0.3$ to $0.7$ with step equal to $0.1$ in SL and WS. Other parameters are the same as those in Fig.~\ref{fig:9}.}
\label{fig:10}
\end{figure}

As mentioned, Eq.~\ref{general transition matrix} describes the general case of game transitions. Therefore, we now turn to game transitions among three states ($G_0$, $G_1$, and $G_2$), where $G_0$ and $G_2$ are the most and least valuable states, respectively, i.e., $b_0 > b_1 > b_2$. In particular, we proceed with the game transition matrices given by Ref.~\cite{su2019evolutionary}, i.e.,
\begin{equation}
\begin{matrix}
\begin{aligned}
\mathbf{P}^{[0]}=\begin{bmatrix}
1-p  & 0 & p\\ 
0  & 1-p & p \\
0 & 0 & 1
\end{bmatrix}
\\
\mathbf{P}^{[1]} =   
\begin{bmatrix}
1-p & p & 0 \\ 
0 & 1 & 0 \\
0 & p & 1-p
\end{bmatrix}
\\
\mathbf{P}^{[2]} = 
\begin{bmatrix}
1 & 0 & 0\\ 
p & 1-p & 0 \\
p & 0 & 1-p
\end{bmatrix}
\end{aligned}
\end{matrix},
\label{threematrices}
\end{equation}
where $0\leq p \leq 1$ is the transition probability. Eq.~\ref{threematrices} indicates that the game transitions from a deterministic to a probabilistic pattern, where mutual cooperation (resp. mutual defection) tends to result in $G_0$ (resp. $G_2$). In contrast, unilateral cooperation or defection yields a moderately valuable $G_1$. For simplicity, we suppose the differences $\Delta b_{i}$ between $G_i$ and $G_{i+1}$ ($i = 0,1$) are the same, denoted as $\Delta b$. In this subsection, we explore how the level of cooperation varies under the joint action of $\Delta b$ and $p$, and typical results are shown in Figs.~\ref{fig:9} and \ref{fig:10}.

In Figs.~\ref{fig:9 1} and \ref{fig:9 2}, we can arrive at the conclusion that the increase in $\Delta b$ and $p$ can significantly promote cooperative behavior. Additionally, since $\Delta b$ in WS is one-half of that set in SL, the escalation in $f_c$ in WS does not exhibit as pronounced as it does in SL. This suggests that the disparity of benefits offered by altruistic actions within games is a crucial factor in the success of cooperation.

Most notably, the unilateral increase of $b_1$ can not facilitate $f_c$ in SL as effectively as WS, as presented in Figs.~\ref{fig:10 1} and \ref{fig:10 2}. Concretely, even if the parameters are set as $b_0 = 10$, $b_1 = 9$, and $b_2 = 8$, $f_c$ is no more than $0.9$, far from the state of pure cooperators, in SL. Conceptually similar phenomena have also been described in the previous section, where only two possible games are considered. 

\subsection{Cooperation Density Affected by Network Size}
\begin{figure}[htbp]
\centering
\subfigure[SL]{
\includegraphics[scale=0.165]{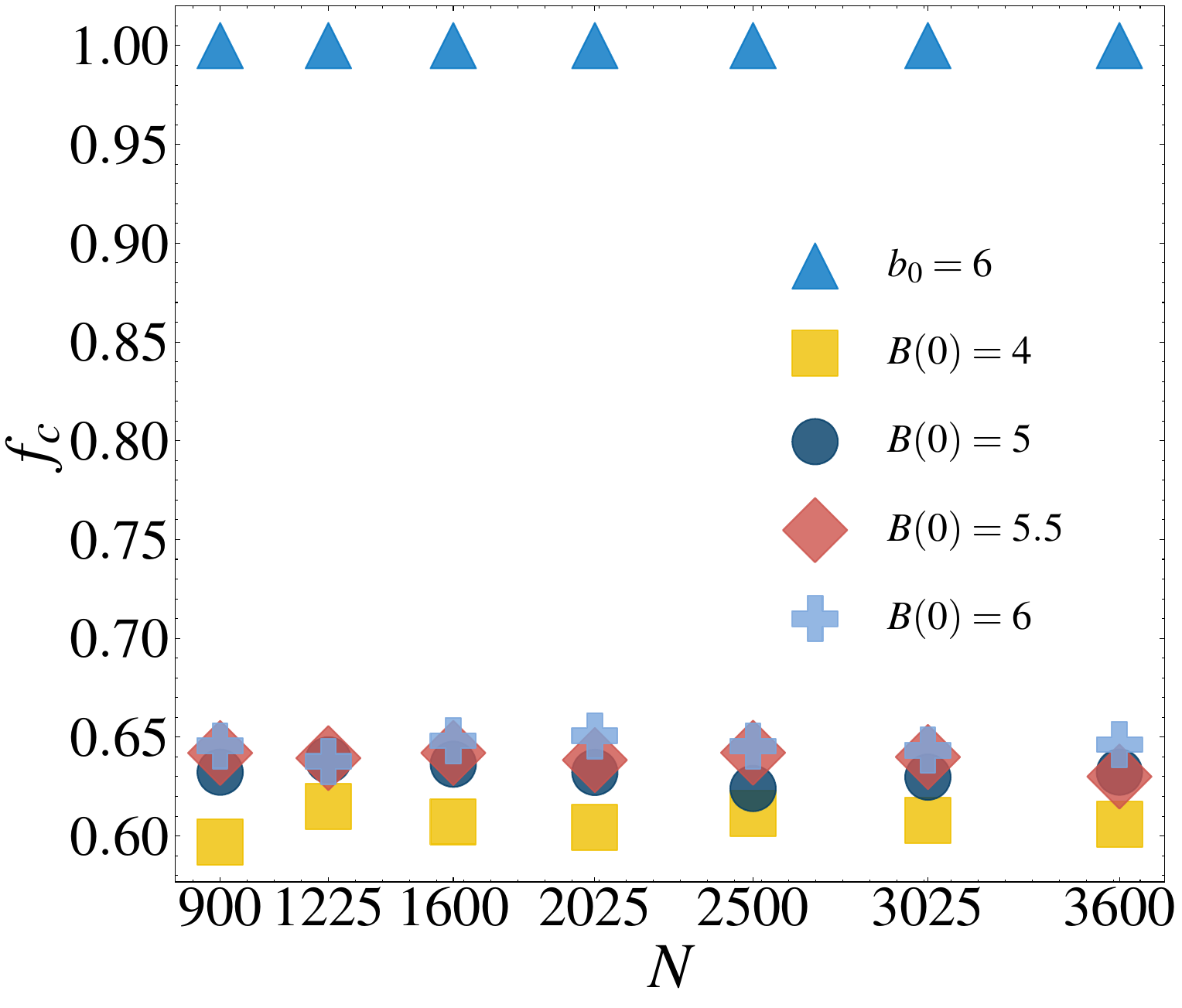}
\label{fig:11 1}
}
\hspace{-5mm}
\subfigure[WS]{
\includegraphics[scale=0.165]{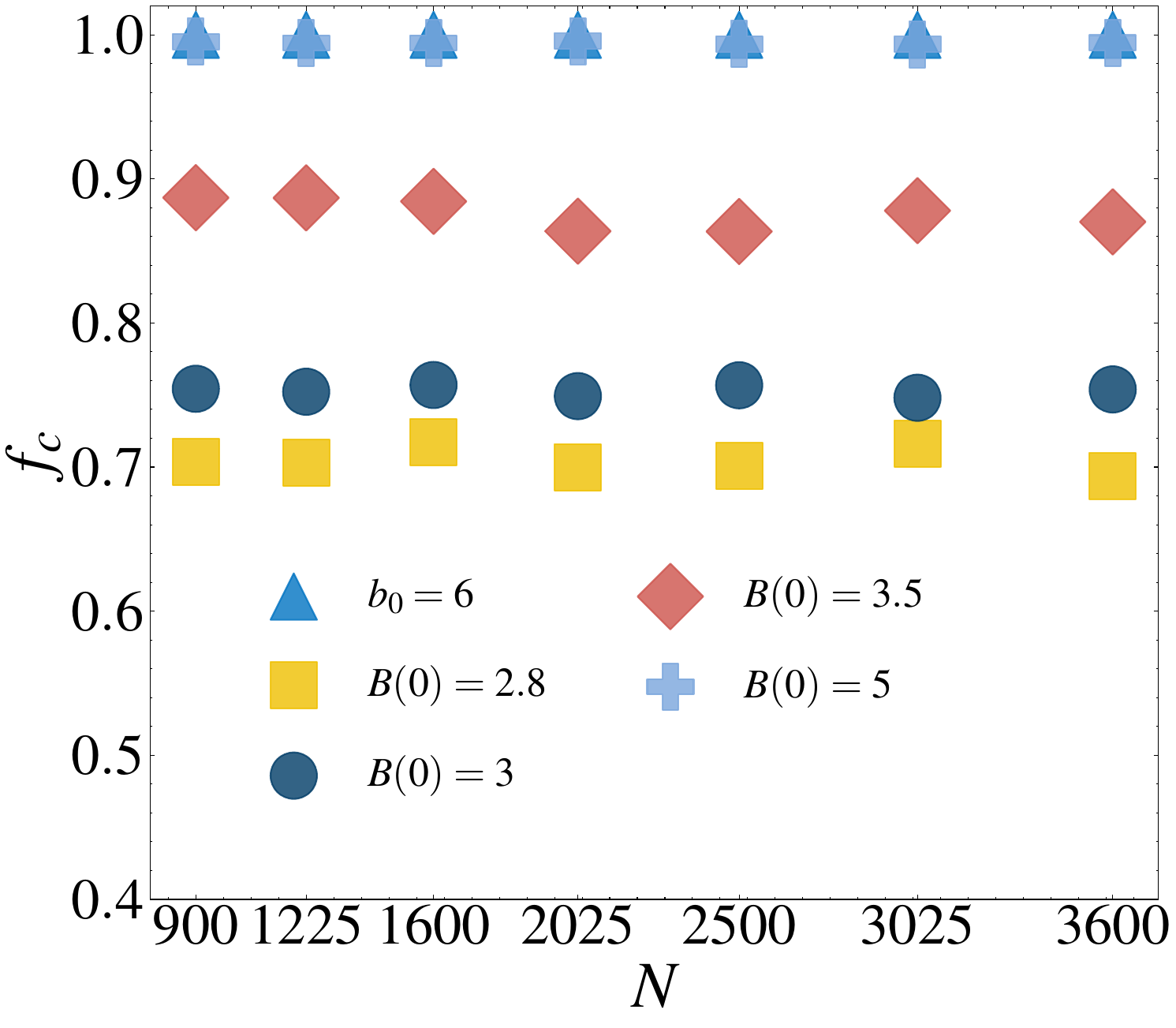}
\label{fig:11 2}
}
\caption{Plots of cooperation density $f_c$ ($y$-axis) against different network scales $N$ ($x$-axis). $N$ varies from the square of $30$ to $60$ with step equal to $5$. The outcomes are averaged over thirty independent simulations to maintain a good accuracy of the simulation results. Other parameters are the same as those in Fig.~\ref{fig:2}.}
\label{fig:11}
\end{figure}

In previous subsections, the network size is fixed to $1600$; hence, it is a crucial question whether the network size affects evolutionary outcomes. To answer this, we here evaluate the robustness of the proposed model by varying network sizes under different pairs of parameters. Representative results in SL and WS are shown in Fig.~\ref{fig:11}, where the network sizes of SL and WS vary from $900$ to $3600$. 

In Figs.~\ref{fig:11 1} and \ref{fig:11 2}, both SL and WS exhibit only minimal fluctuation in $f_c$. More precisely, in SL, the variances for the five cases, from top
to bottom corresponding to the order of the legend, are $(0.0, 2.8, 1.9, 1.6, 1.4) \times 10^{-5}$ respectively, and similarly in WS, from top to bottom, the corresponding variances are $(0.0, 6.4, 1.0, 9.3, 0.051) \times 10^{-5}$. The results suggest that the network scale has a negligible impact on cooperative behavior when the network scale is substantial, highlighting the robustness of our model behavior. Moreover, for PBIGTs in SL, though under different sizes, $f_c$ does not show explicit differences, which aligns with our previous analysis. 

\subsection{Evolution of Cooperation with Mutation}
\begin{figure}[htbp]
\centering
\subfigure[SL]{
\includegraphics[scale=0.165]{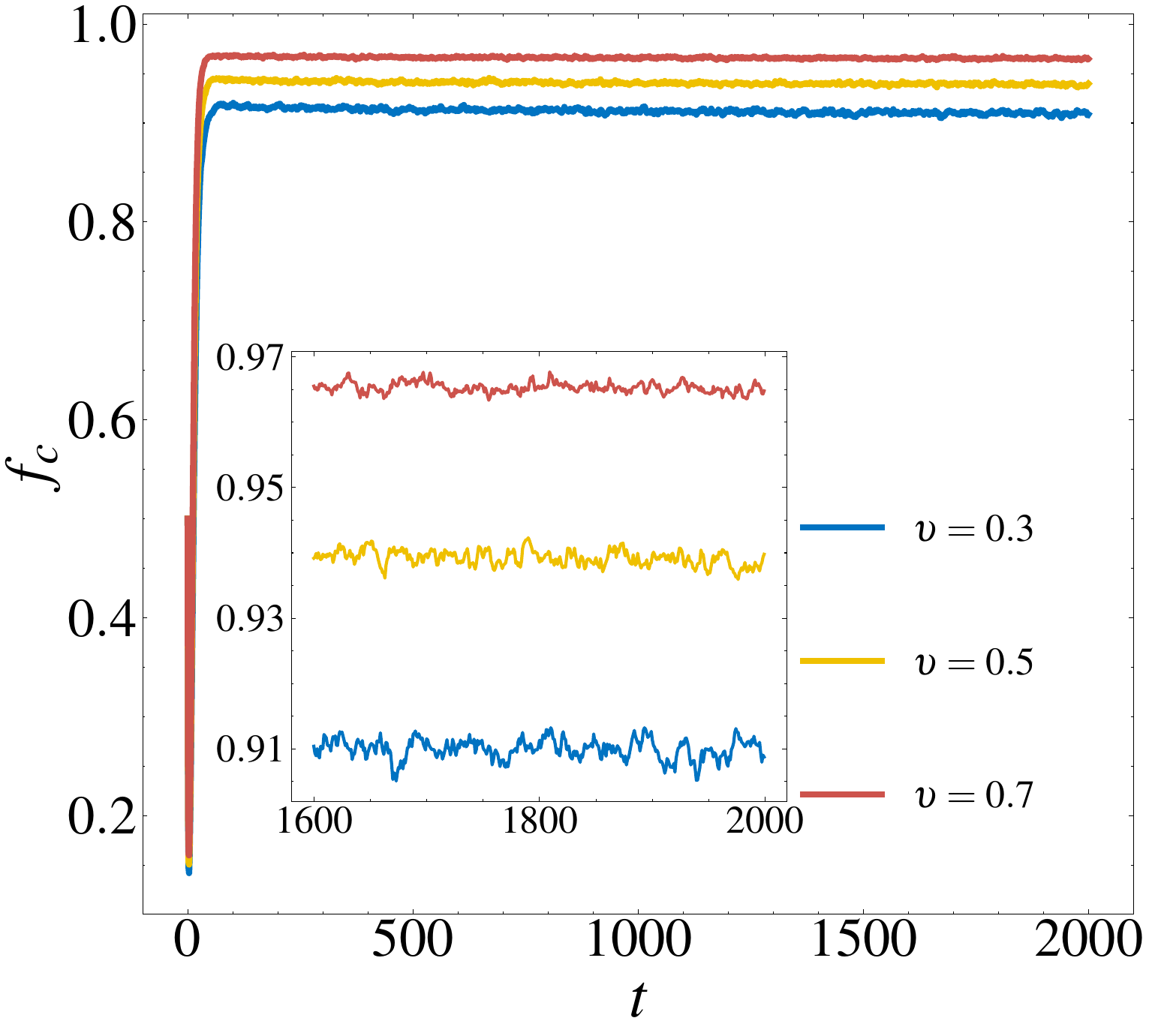}
\label{fig:12 1}
}
\hspace{-5mm}
\subfigure[WS]{
\includegraphics[scale=0.165]{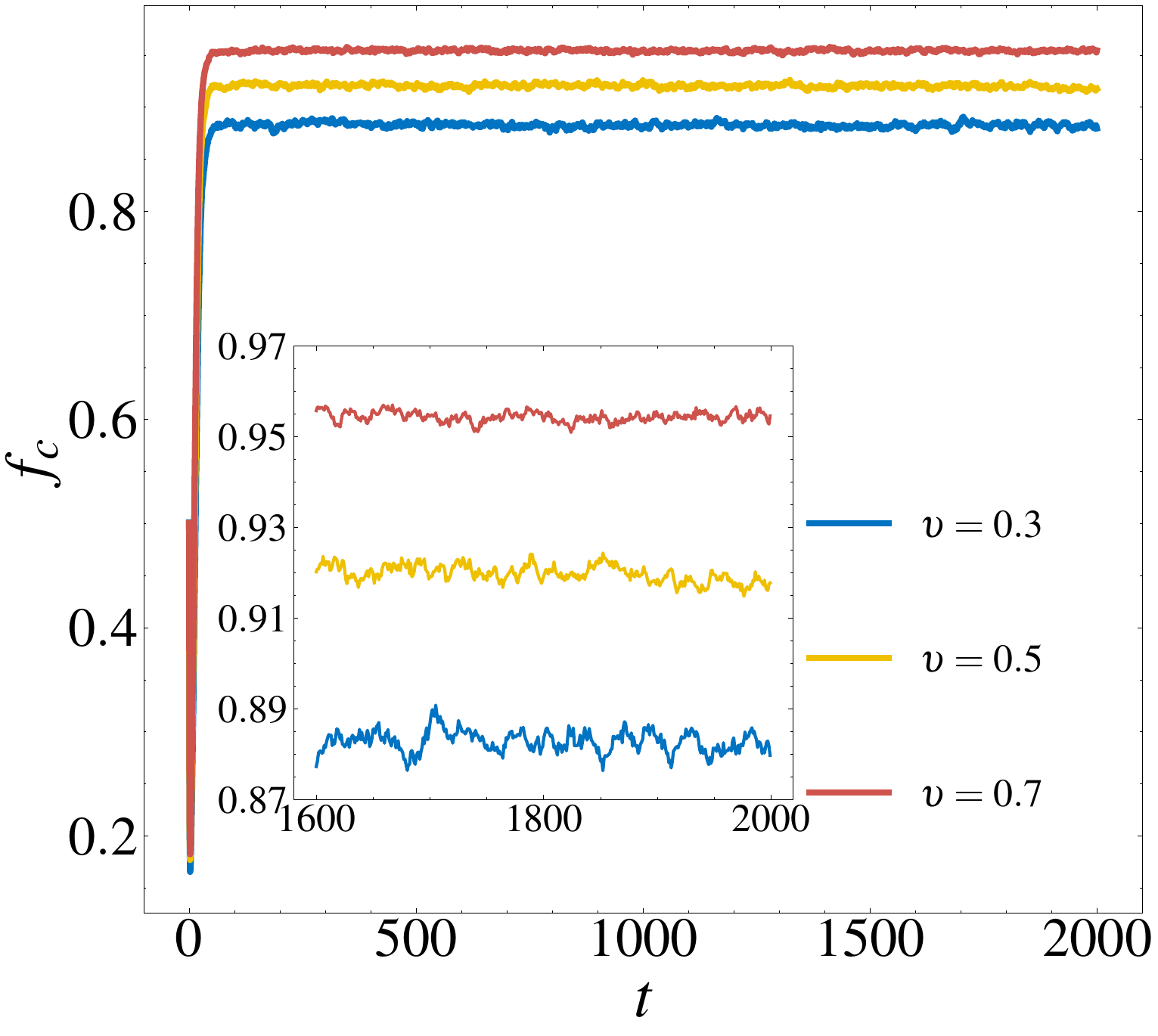}
\label{fig:12 2}
}
\caption{Plots of cooperation frequency against evolutionary time with mutation. This figure shows the evolution of $f_c$ over time on SL (in subplot (a)) and WS (in subplot (b)), respectively, where the $x$-axis is the time $t$ and $y$-axis corresponds to the cooperation frequency $f_c$, under parameters $b_0 = 4.5$, $b_1=3$, and $\mu=0.02$. Other parameters are: $c=1$, $\delta=0.2$, $\kappa=1$, and $a=4$.}
\label{fig:12}
\end{figure}

\begin{figure}[htbp]
\centering
\subfigure[SL]{
\includegraphics[scale=0.17]{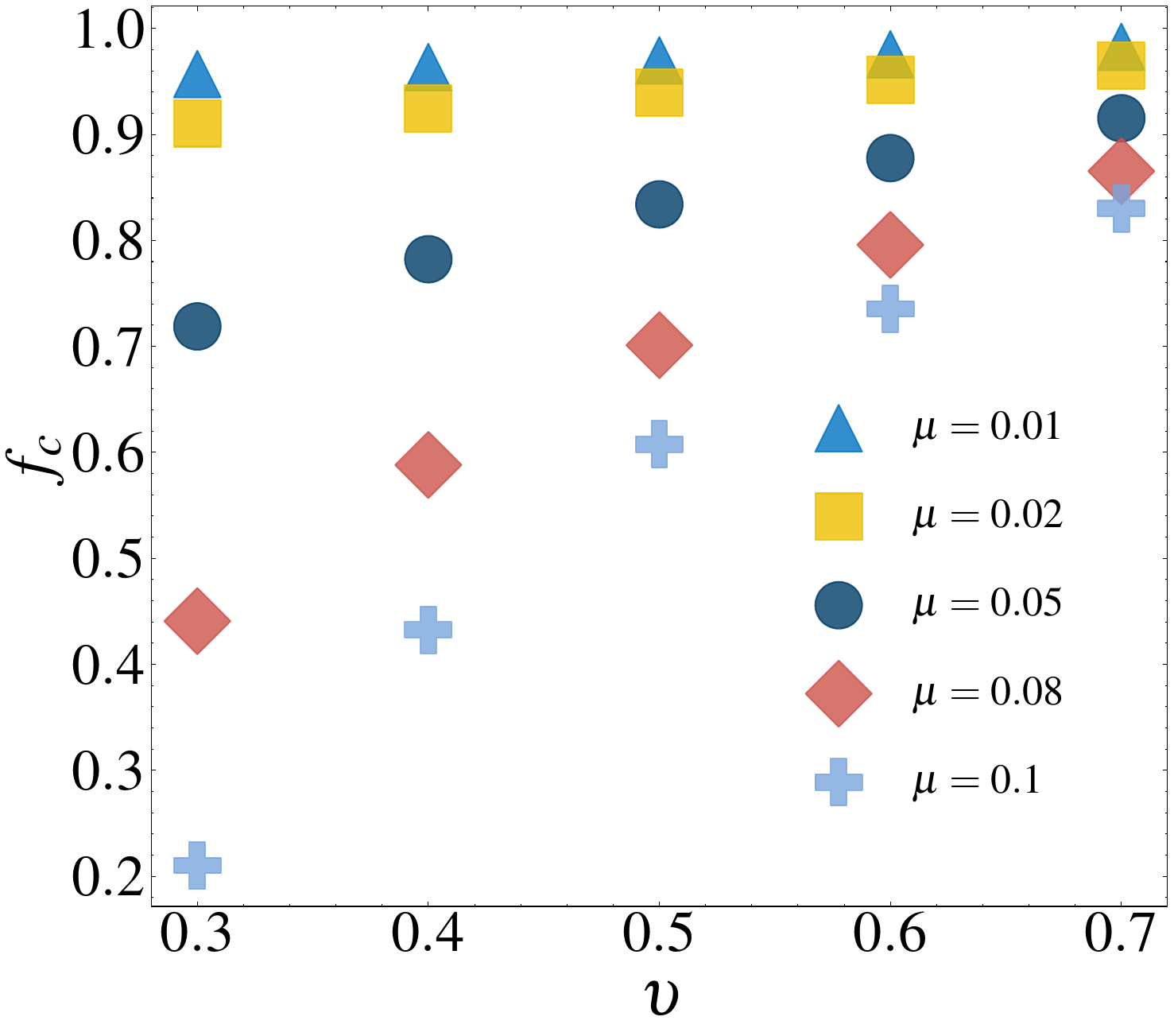}
\label{fig:13 1}
}
\hspace{-5mm}
\subfigure[WS]{
\includegraphics[scale=0.17]{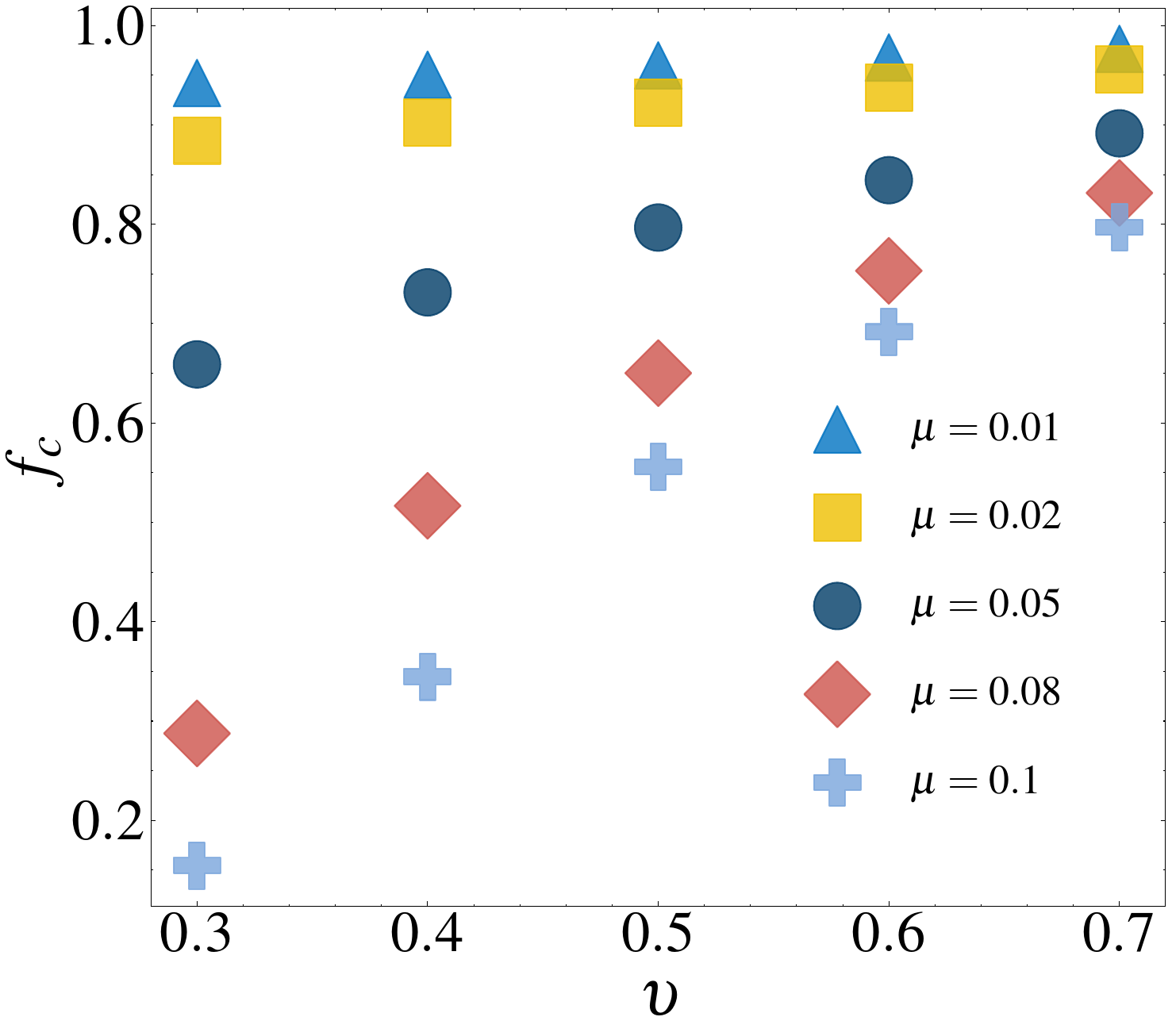}
\label{fig:13 2}
}
\caption{Plots of cooperation frequency against different mutational biases. We present $f_c$ ($y$-axis) against mutational biases $\upsilon \in \{0.3,0.4,0.5,0.6,0.7\}$ ($x$-axis) on SL (in panel (a)) and WS (in panel (b)) under parameters $b_0 = 4.5$ and $b_1=3$. We set $\mu \in \{0.01,0.02,0.05,0.08,0.1\}$. Other parameters are the same as those in Fig.~\ref{fig:12}.}
\label{fig:13}
\end{figure}

In this subsection, considering mutation (or random strategy exploration \cite{traulsen2009exploration}) plays an important role in understanding the emergence of cooperation. Hence, we relax the assumption that the strategy updating of individuals consistently adheres to the revised Fermi function described in Eq.~\ref{fermi function}. Therefore, we introduce the mutation rate (or exploration rate) $0\leq \mu \leq 1$ (although the same mathematical notation $\mu$ as in Eq.~\ref{GBM} is used, no ambiguity will arise in the specific context). More specifically, with probability $1-\mu$, a strategy update goes with the Fermi function. Otherwise, a random mutation occurs (with probability $\mu$), with a probability of $\upsilon$ to $C$ and $1-\upsilon$ to $D$, where $\upsilon$ is known as the mutational bias \cite{allen2019mathematical}. In conventional scenarios, $\upsilon$ is set as $1/2$ by default, i.e., the mutation is unbiased. Therefore, the update of $i$'s strategy at $t$ can be expressed as follows:
\begin{equation}
    s_{i}^{t+1}=\begin{cases}
	s_{\beta(i)}^{t},\mathrm{with}\ \mathrm{probability}\ (1-\mu)\mathbb{P}( s_i\gets s_{\beta(i)}), \\
	s_{i}^{t},\mathrm{with}\ \mathrm{probability}\ (1-\mu)[1-\mathbb{P}(s_i\gets s_{\beta(i)})],\\
	C,\mathrm{with}\ \mathrm{probability}\ \mu\upsilon,\\
        D,\mathrm{with}\ \mathrm{probability}\ \mu(1-\upsilon).\\
\end{cases}
\end{equation}
In Figs.~\ref{fig:12} and \ref{fig:13}, we demonstrate how the variation of $\mu$ and $\upsilon$ affect $f_c$. 

In Fig.~\ref{fig:12}, we exhibit how $f_c$ varies with the evolutionary time $t$. The results indicate that game transitions can effectively ensure the emergence of cooperation, even in the presence of mutation. Furthermore, as $\upsilon$ increases, the higher and more stable the level of cooperation becomes. No matter in SL and WS, $f_c$ finally reaches one stationary distribution. To further explore the influence of $\mu$ and $\upsilon$ on $f_c$, typical results are shown in Figs.~\ref{fig:13 1} and \ref{fig:13 2}.

In Fig.~\ref{fig:13}, as we can clearly see, for relatively large $\mu$ and relatively small $\upsilon$, cooperation collapses because the high-frequency emergence of a mutant defector within a cluster of cooperators spoils the spatial assortment of cooperative individuals and thus, inhibits the emergence of cooperation \cite{allen2012mutation}. Moreover, in other cases, the evolving system would finally enter a stationary distribution, as the ones shown in Figs.~\ref{fig:12 1} and \ref{fig:12 2}. Particularly, in the proposed model, it can be regarded as one special stationary distribution under the combined effect of game transitions, reputations, mutations, and selections, similar to the well-known ``mutation-selection'' stationary distribution \cite{allen2019mathematical,allen2014measures} and ``game-mutation-selection'' stationary distribution \cite{su2019evolutionary}.

To summarize our findings, we can conclude that the proposed mechanism can make cooperation stable even when a random strategy exploration is considered. It is found that when the mutation rate $\mu$ is relatively small and the mutational bias $\upsilon$ is relatively large, cooperation can still dominate the networked population over defection.

\section{Conclusion and Discussion}
The primary goal of this study was to elucidate the impact of reputation on cooperative behavior in temporally evolving environments, which stem from both intrinsic and extrinsic dynamics. To obtain a comprehensive view of the potential system behavior, we have concentrated on two significant topologies, i.e., regular square-lattice and highly random Watts–Strogatz small-world networks. Our work is somewhat different compared to the existing literature, which assumes a direct increase in the reputation of focal cooperators. More specifically, our work proposed a growth mechanism of reputation reflected by the cooperation level among its neighbors, which is determined by the number of cooperators. It is recognized that cooperation survives by forming clusters to mitigate the exploitation by defectors \cite{hauert2004spatial}. Therefore, it is unambiguous that such a mechanism is beneficial for cooperators when clusters are formed, and it is unfavorable to cooperation before formation. It is the most different from the previous mechanism, where in both mentioned cases, the reputation of cooperators increases more than that of defectors, as a direct reward for the altruistic behavior. By comprehensive and intensive simulations, we have uncovered that increasing the reputation strength $\delta$ during the imitation process benefits the cooperative behavior, but the reputation-fitness parameter $a$ does not work significantly. Additionally, an interesting phenomenon has been observed regarding the influence of game transition, where the unilateral value of $b_i$ fails to turn the whole networked population into a pure cooperator state in SL, indicating that the difference between games matters. 

To conduct a thorough investigation, the effects of biased mutation are taken into account. We observed that the cooperation frequency evolves into one ``selection-mutation-reputation-game'' stationary distribution under specific parameters: the mutation rate $\mu$ and the mutational bias $\upsilon$. When the mutation rate is relatively large and the mutational bias is relatively small, the spatial assortment of cooperators would be diluted, thus resulting in the dominance of defection.

Furthermore, the robustness of the proposed model is examined from two aspects. On the one hand, concerning whether the initial game condition would affect strategy evolution, we test different game distributions in the initial state and discuss a special case where the game transition degenerates into a static one. On the other hand, to assess the influence of network size on the prevalence of cooperation, we implement a broad range of scales in our analysis. 

It is worth noting that, in recent and related works, the changing environments exert distinctly disparate impacts on evolutionary dynamics. For example, Stewart and Plotkin demonstrated that in iterated two-player games, a collapse of cooperation arises when there is a tradeoff between the benefits and costs of cooperation \cite{stewart2014collapse}. The persistent oscillations of strategies were observed with the ``game-environment'' feedback in Refs.~\cite{weitz2016oscillating,liu_lj_rsif22}. In their study, Assaf {\it et al.} illustrated that extrinsic noise related to the selection strength, modeled by an Ornstein-Uhlenbeck process, has the potential to improve the probability of cooperation fixation greatly \cite{assaf2013cooperation}. Szolnoki and Chen detected that the proposed ``cooperation-environment'' coevolutionary rule in Ref.~\cite{szolnoki2018environmental} significantly fertilized the flourish of cooperation. In comparison to recent works in a changing environment, our model exhibits a degree of distinction, accounting for both extrinsic factors and individuals' behaviors that facilitate a heterogeneous change of the environment under the reputation mechanism.  

This study mainly discusses transitions among different dilemma strengths determined by individuals' mutual behaviors to model the partner-fidelity feedback phenomenon widely existing in evolutionary biology \cite{sachs2004evolution,wulff2008life}. Meanwhile, individuals may encounter distinct game types across various social contexts in reality, such as the battle of the sexes, hawk-dove game, and chicken game. Even further, the combined case, where both game mode and its corresponding strength are allowed to vary according to certain reasonable rules, is all the more attractive and intriguing. Therefore, further exploration is an important goal and pursuit for future work. Ultimately, we aspire for our investigation to make a meaningful contribution to the realm of evolutionary game theory, specifically concerning reputation mechanisms, game transitions, and associated elements, thereby augmenting the practical applicability of these subjects \cite{li2023open}.

\section*{Acknowledgment}
This work was partly supported by the National Natural Science Foundation of China (NSFC) under Grant No. 62206230, the Natural Science Foundation of Chongqing under Grant No. CSTB2023NSCQ-MSX0064, and the National Research, Development and Innovation Office (NKFIH) under Grant No. K142948.

\ifCLASSOPTIONcaptionsoff
  \newpage
\fi



%

\bibliographystyle{IEEEtran}
\bibliography{cas-refs}

@article{su2019evolutionary,
  title={Evolutionary dynamics with game transitions},
  author={Su, Qi and McAvoy, Alex and Wang, Long and Nowak, Martin A},
  journal={Proceedings of the National Academy of Sciences},
  volume={116},
  number={51},
  pages={25398--25404},
  year={2019},
  publisher={National Acad Sciences}
}

@article{allen2017evolutionary,
  title={Evolutionary dynamics on any population structure},
  author={Allen, Benjamin and Lippner, Gabor and Chen, Yu-Ting and Fotouhi, Babak and Momeni, Naghmeh and Yau, Shing-Tung and Nowak, Martin A},
  journal={Nature},
  volume={544},
  number={7649},
  pages={227--230},
  year={2017},
  publisher={Nature Publishing Group UK London}
}

@article{dafoe2021cooperative,
  title={Cooperative \protect{AI}: machines must learn to find common ground},
  author={Dafoe, Allan and Bachrach, Yoram and Hadfield, Gillian and Horvitz, Eric and Larson, Kate and Graepel, Thore},
  journal={Nature},
  volume={593},
  number={7857},
  pages={33--36},
  year={2021},
  publisher={Nature Publishing Group UK London}
}

@book{darwin2004origin,
  title={On the origin of species, 1859},
  author={Darwin, Charles},
  year={2004},
  publisher={Routledge}
}

@article{feng2023evolutionary,
  title={An evolutionary game with the game transitions based on the \protect{Markov} process},
  author={Feng, Minyu and Pi, Bin and Deng, Liang-Jian and Kurths, J{\"u}rgen},
  journal={IEEE Transactions on Systems, Man, and Cybernetics: Systems},
  volume={54},
  number={1},
  pages={609--621},
  year={2024},
  publisher={IEEE}
}

@article{hilbe2018evolution,
  title={Evolution of cooperation in stochastic games},
  author={Hilbe, Christian and {\v{S}}imsa, {\v{S}}t{\v{e}}p{\'a}n and Chatterjee, Krishnendu and Nowak, Martin A},
  journal={Nature},
  volume={559},
  number={7713},
  pages={246--249},
  year={2018},
  publisher={Nature Publishing Group UK London}
}

@article{nowak1992evolutionary,
  title={Evolutionary games and spatial chaos},
  author={Nowak, Martin A and May, Robert M},
  journal={Nature},
  volume={359},
  number={6398},
  pages={826--829},
  year={1992},
  publisher={Nature Publishing Group UK London}
}

@article{watts1998collective,
  title={Collective dynamics of ‘small-world’networks},
  author={Watts, Duncan J and Strogatz, Steven H},
  journal={Nature},
  volume={393},
  number={6684},
  pages={440--442},
  year={1998},
  publisher={Nature Publishing Group}
}

@article{barabasi1999emergence,
  title={Emergence of scaling in random networks},
  author={Barab{\'a}si, Albert-L{\'a}szl{\'o} and Albert, R{\'e}ka},
  journal={Science},
  volume={286},
  number={5439},
  pages={509--512},
  year={1999},
  publisher={American Association for the Advancement of Science}
}

@article{nowak2006five,
  title={Five rules for the evolution of cooperation},
  author={Nowak, Martin A},
  journal={Science},
  volume={314},
  number={5805},
  pages={1560--1563},
  year={2006},
  publisher={American Association for the Advancement of Science}
}

@article{allen2015molecular,
  title={The molecular clock of neutral evolution can be accelerated or slowed by asymmetric spatial structure},
  author={Allen, Benjamin and Sample, Christine and Dementieva, Yulia and Medeiros, Ruben C and Paoletti, Christopher and Nowak, Martin A},
  journal={PLoS Computational Biology},
  volume={11},
  number={2},
  pages={e1004108},
  year={2015},
  publisher={Public Library of Science San Francisco, CA USA}
}

@article{ohtsuki2006simple,
  title={A simple rule for the evolution of cooperation on graphs and social networks},
  author={Ohtsuki, Hisashi and Hauert, Christoph and Lieberman, Erez and Nowak, Martin A},
  journal={Nature},
  volume={441},
  number={7092},
  pages={502--505},
  year={2006},
  publisher={Nature Publishing Group UK London}
}

@article{santos2005scale,
  title={Scale-free networks provide a unifying framework for the emergence of cooperation},
  author={Santos, Francisco C and Pacheco, Jorge M},
  journal={Physical Review Letters},
  volume={95},
  number={9},
  pages={098104},
  year={2005},
  publisher={APS}
}

@article{yao2023inhibition,
  title={Inhibition and activation of interactions in networked weak prisoner’s dilemma},
  author={Yao, Yichao and Zeng, Ziyan and Pi, Bin and Feng, Minyu},
  journal={Chaos: An Interdisciplinary Journal of Nonlinear Science},
  volume={33},
  number={6},
  pages={063124},
  year={2023},
  publisher={AIP Publishing}
}

@article{allen2019mathematical,
  title={A mathematical formalism for natural selection with arbitrary spatial and genetic structure},
  author={Allen, Benjamin and McAvoy, Alex},
  journal={Journal of Mathematical Biology},
  volume={78},
  number={4},
  pages={1147--1210},
  year={2019},
  publisher={Springer}
}

@article{allen2014measures,
  title={Measures of success in a class of evolutionary models with fixed population size and structure},
  author={Allen, Benjamin and Tarnita, Corina E},
  journal={Journal of Mathematical Biology},
  volume={68},
  pages={109--143},
  year={2014},
  publisher={Springer}
}

@article{allen2012mutation,
  title={How mutation affects evolutionary games on graphs},
  author={Allen, Benjamin and Traulsen, Arne and Tarnita, Corina E and Nowak, Martin A},
  journal={Journal of Theoretical Biology},
  volume={299},
  pages={97--105},
  year={2012},
  publisher={Elsevier}
}

@article{su2019edgediversity,
  title={Evolutionary multiplayer games on graphs with edge diversity},
  author={Su, Qi and Zhou, Lei and Wang, Long},
  journal={PLoS Computational Biology},
  volume={15},
  number={4},
  pages={e1006947},
  year={2019},
  publisher={Public Library of Science San Francisco, CA USA}
}

@article{traulsen2009exploration,
  title={Exploration dynamics in evolutionary games},
  author={Traulsen, Arne and Hauert, Christoph and De Silva, Hannelore and Nowak, Martin A and Sigmund, Karl},
  journal={Proceedings of the National Academy of Sciences},
  volume={106},
  number={3},
  pages={709--712},
  year={2009},
  publisher={National Acad Sciences}
}

@article{perc2010coevolutionary,
  title={Coevolutionary games—a mini review},
  author={Perc, Matja{\v{z}} and Szolnoki, Attila},
  journal={BioSystems},
  volume={99},
  number={2},
  pages={109--125},
  year={2010},
  publisher={Elsevier}
}

@article{majhi2022dynamics,
  title={Dynamics on higher-order networks: A review},
  author={Majhi, Soumen and Perc, Matja{\v{z}} and Ghosh, Dibakar},
  journal={Journal of the Royal Society Interface},
  volume={19},
  number={188},
  pages={20220043},
  year={2022},
  publisher={The Royal Society}
}

@article{xiong2024coevolution,
  title={Coevolution of relationship and interaction in cooperative dynamical multiplex networks},
  author={Xiong, Xiaojin and Zeng, Ziyan and Feng, Minyu and Szolnoki, Attila},
  journal={Chaos: An Interdisciplinary Journal of Nonlinear Science},
  volume={34},
  number={2},
  pages={023118},
  year={2024},
  publisher={AIP Publishing}
}

@article{boccaletti2014structure,
  title={The structure and dynamics of multilayer networks},
  author={Boccaletti, Stefano and Bianconi, Ginestra and Criado, Regino and Del Genio, Charo I and G{\'o}mez-Gardenes, Jes{\'u}s and Romance, Miguel and Sendina-Nadal, Irene and Wang, Zhen and Zanin, Massimiliano},
  journal={Physics Reports},
  volume={544},
  number={1},
  pages={1--122},
  year={2014},
  publisher={Elsevier}
}

@article{alvarez2021evolutionary,
  title={Evolutionary dynamics of higher-order interactions in social networks},
  author={Alvarez-Rodriguez, Unai and Battiston, Federico and de Arruda, Guilherme Ferraz and Moreno, Yamir and Perc, Matja{\v{z}} and Latora, Vito},
  journal={Nature Human Behaviour},
  volume={5},
  number={5},
  pages={586--595},
  year={2021},
  publisher={Nature Publishing Group UK London}
}

@article{szolnoki2009impact,
  title={Impact of aging on the evolution of cooperation in the spatial prisoner’s dilemma game},
  author={Szolnoki, Attila and Perc, Matja{\v{z}} and Szab{\'o}, Gy{\"o}rgy and Stark, Hans-Ulrich},
  journal={Physical Review E},
  volume={80},
  number={2},
  pages={021901},
  year={2009},
  publisher={APS}
}

@article{szolnoki2008coevolution,
  title={Coevolution of teaching activity promotes cooperation},
  author={Szolnoki, Attila and Perc, Matja{\v{z}}},
  journal={New Journal of Physics},
  volume={10},
  number={4},
  pages={043036},
  year={2008},
  publisher={IOP Publishing}
}

@article{xiao2020leaving,
  title={Leaving bads provides better outcome than approaching goods in a social dilemma},
  author={Xiao, Zhilong and Chen, Xiaojie and Szolnoki, Attila},
  journal={New Journal of Physics},
  volume={22},
  number={2},
  pages={023012},
  year={2020},
  publisher={IOP Publishing}
}

@article{rankin2007tragedy,
  title={The tragedy of the commons in evolutionary biology},
  author={Rankin, Daniel J and Bargum, Katja and Kokko, Hanna},
  journal={Trends in Ecology \& Evolution},
  volume={22},
  number={12},
  pages={643--651},
  year={2007},
  publisher={Elsevier}
}

@article{mcfall2013animals,
  title={Animals in a bacterial world, a new imperative for the life sciences},
  author={McFall-Ngai, Margaret and Hadfield, Michael G and Bosch, Thomas CG and Carey, Hannah V and Domazet-Lo{\v{s}}o, Tomislav and Douglas, Angela E and Dubilier, Nicole and Eberl, Gerard and Fukami, Tadashi and Gilbert, Scott F and others},
  journal={Proceedings of the National Academy of Sciences},
  volume={110},
  number={9},
  pages={3229--3236},
  year={2013},
  publisher={National Acad Sciences}
}

@book{sigmund2010calculus,
  title={The calculus of selfishness},
  author={Sigmund, Karl},
  year={2010},
  publisher={Princeton University Press}
}

@article{sachs2004evolution,
  title={The evolution of cooperation},
  author={Sachs, Joel L and Mueller, Ulrich G and Wilcox, Thomas P and Bull, James J},
  journal={The Quarterly Review of Biology},
  volume={79},
  number={2},
  pages={135--160},
  year={2004},
  publisher={The University of Chicago Press}
}

@article{su2022evolution,
  title={Evolution of cooperation with contextualized behavior},
  author={Su, Qi and McAvoy, Alex and Plotkin, Joshua B},
  journal={Science Advances},
  volume={8},
  number={6},
  pages={eabm6066},
  year={2022},
  publisher={American Association for the Advancement of Science}
}

@article{xia2023reputation,
  title={Reputation and reciprocity},
  author={Xia, Chengyi and Wang, Juan and Perc, Matja{\v{z}} and Wang, Zhen},
  journal={Physics of Life Reviews},
  year={2023},
  publisher={Elsevier}
}

@article{nowak1998evolution,
  title={Evolution of indirect reciprocity by image scoring},
  author={Nowak, Martin A and Sigmund, Karl},
  journal={Nature},
  volume={393},
  number={6685},
  pages={573--577},
  year={1998},
  publisher={Nature Publishing Group UK London}
}

@article{panchanathan2003tale,
  title={A tale of two defectors: the importance of standing for evolution of indirect reciprocity},
  author={Panchanathan, Karthik and Boyd, Robert},
  journal={Journal of Theoretical Biology},
  volume={224},
  number={1},
  pages={115--126},
  year={2003},
  publisher={Elsevier}
}

@article{erovenko2019effect,
  title={The effect of network topology on optimal exploration strategies and the evolution of cooperation in a mobile population},
  author={Erovenko, Igor V and Bauer, Johann and Broom, Mark and Pattni, Karan and Rycht{\'a}{\v{r}}, Jan},
  journal={Proceedings of the Royal Society A},
  volume={475},
  number={2230},
  pages={20190399},
  year={2019},
  publisher={The Royal Society Publishing}
}

@article{mcpherson2001birds,
  title={Birds of a feather: Homophily in social networks},
  author={McPherson, Miller and Smith-Lovin, Lynn and Cook, James M},
  journal={Annual Review of Sociology},
  volume={27},
  number={1},
  pages={415--444},
  year={2001},
  publisher={Annual Reviews 4139 El Camino Way, PO Box 10139, Palo Alto, CA 94303-0139, USA}
}

@article{zhang2022reputation,
  title={Reputation-based asymmetric comparison of fitness promotes cooperation on complex networks},
  author={Zhang, Zhipeng and Wu, Yu’e and Zhang, Shuhua},
  journal={Physica A: Statistical Mechanics and its Applications},
  volume={608},
  pages={128268},
  year={2022},
  publisher={Elsevier}
}

@article{santos2016evolution,
  title={Evolution of cooperation under indirect reciprocity and arbitrary exploration rates},
  author={Santos, Fernando P and Pacheco, Jorge M and Santos, Francisco C},
  journal={Scientific Reports},
  volume={6},
  number={1},
  pages={37517},
  year={2016},
  publisher={Nature Publishing Group UK London}
}

@article{weitz2016oscillating,
  title={An oscillating tragedy of the commons in replicator dynamics with game-environment feedback},
  author={Weitz, Joshua S and Eksin, Ceyhun and Paarporn, Keith and Brown, Sam P and Ratcliff, William C},
  journal={Proceedings of the National Academy of Sciences},
  volume={113},
  number={47},
  pages={E7518--E7525},
  year={2016},
  publisher={National Acad Sciences}
}

@article{szolnoki2018environmental,
  title={Environmental feedback drives cooperation in spatial social dilemmas},
  author={Szolnoki, Attila and Chen, Xiaojie},
  journal={Europhysics Letters},
  volume={120},
  number={5},
  pages={58001},
  year={2018},
  publisher={IOP Publishing}
}

@article{wulff2008life,
  title={Life-history differences among coral reef sponges promote mutualism or exploitation of mutualism by influencing partner fidelity feedback},
  author={Wulff, Janie L},
  journal={The American Naturalist},
  volume={171},
  number={5},
  pages={597--609},
  year={2008},
  publisher={The University of Chicago Press}
}

@article{assaf2013cooperation,
  title={Cooperation dilemma in finite populations under fluctuating environments},
  author={Assaf, Michael and Mobilia, Mauro and Roberts, Elijah},
  journal={Physical Review Letters},
  volume={111},
  number={23},
  pages={238101},
  year={2013},
  publisher={APS}
}

@article{stewart2014collapse,
  title={Collapse of cooperation in evolving games},
  author={Stewart, Alexander J and Plotkin, Joshua B},
  journal={Proceedings of the National Academy of Sciences},
  volume={111},
  number={49},
  pages={17558--17563},
  year={2014},
  publisher={National Acad Sciences}
}

@ARTICLE{ liu_lj_rsif22,
    AUTHOR  = {Linjie Liu and Zhilong Xiao and Xiaojie Chen and Attila Szolnoki},
    TITLE   = {Early exclusion leads to cyclical cooperation in repeated group interactions},
    JOURNAL = {J. R. Soc. Interface},
    YEAR    = {2022},
    VOLUME  = {19},
    PAGES   = {20210755},
}

@inproceedings{zhang2021incentive,
  title={Incentive mechanism for horizontal federated learning based on reputation and reverse auction},
  author={Zhang, Jingwen and Wu, Yuezhou and Pan, Rong},
  booktitle={Proceedings of the Web Conference 2021},
  pages={947--956},
  year={2021}
}

@inproceedings{gkorou2012reducing,
  title={Reducing the history in decentralized interaction-based reputation systems},
  author={Gkorou, Dimitra and Vink{\'o}, Tam{\'a}s and Chiluka, Nitin and Pouwelse, Johan and Epema, Dick},
  booktitle={NETWORKING 2012: 11th International IFIP TC 6 Networking Conference, Prague, Czech Republic, May 21-25, 2012, Proceedings, Part II 11},
  pages={238--251},
  year={2012},
  organization={Springer}
}

@article{li2019reputation,
  title={Reputation-based adaptive adjustment of link weight among individuals promotes the cooperation in spatial social dilemmas},
  author={Li, Xiaopeng and Sun, Shiwen and Xia, Chengyi},
  journal={Applied Mathematics and Computation},
  volume={361},
  pages={810--820},
  year={2019},
  publisher={Elsevier}
}

@article{mao2021effect,
  title={Effect of collective influence on the evolution of cooperation in evolutionary prisoner’s dilemma games},
  author={Mao, Yajun and Rong, Zhihai and Wu, Zhi-Xi},
  journal={Applied Mathematics and Computation},
  volume={392},
  pages={125679},
  year={2021},
  publisher={Elsevier}
}

@article{li2023open,
  author={Li, Qin and Pi, Bin and Feng, Minyu and Kurths, Jürgen},
  journal={IEEE Transactions on Computational Social Systems}, 
  title={Open Data in the Digital Economy: An Evolutionary Game Theory Perspective}, 
  year={2024},
  volume={11},
  number={3},
  pages={3780-3791}
}

@ARTICLE{qu2022evolutionary,
  author={Qu, Cunquan and Ji, Chenlu and Zhang, Ming},
  journal={IEEE Transactions on Computational Social Systems}, 
  title={Evolutionary Game Model With Group Decision-Making in Signed Social Networks}, 
  year={2023},
  volume={10},
  number={6},
  pages={3444-3453}
}

@article{hauert2004spatial,
  title={Spatial structure often inhibits the evolution of cooperation in the snowdrift game},
  author={Hauert, Christoph and Doebeli, Michael},
  journal={Nature},
  volume={428},
  number={6983},
  pages={643--646},
  year={2004},
  publisher={Nature Publishing Group UK London}
}

@article{li2022graphical,
  author={Li, Yuejiang and Zhao, H. Vicky and Chen, Yan},
  journal={IEEE Transactions on Signal and Information Processing over Networks}, 
  title={Graphical Evolutionary Game Theoretic Modeling of Strategy Evolution Over Heterogeneous Networks}, 
  year={2022},
  volume={8},
  number={},
  pages={739-754},
  doi={10.1109/TSIPN.2022.3202308}
}

@article{banez2022modeling,
  author={Banez, Reginald A. and Gao, Hao and Li, Lixin and Yang, Chungang and Han, Zhu and Poor, H. Vincent},
  journal={IEEE Transactions on Signal and Information Processing over Networks}, 
  title={Modeling and Analysis of Opinion Dynamics in Social Networks Using Multiple-Population Mean Field Games}, 
  year={2022},
  volume={8},
  number={},
  pages={301-316},
  doi={10.1109/TSIPN.2022.3166102}
}

@article{zhou2024coevolution,
  title={Coevolution of extortion strategies with mixed imitation and aspiration learning dynamics in spatial Prisoner’s Dilemma game},
  author={Zhou, Zhizhuo and Rong, Zhihai and Yang, Wen and Wu, Zhi-Xi},
  journal={Chaos, Solitons \& Fractals},
  volume={188},
  pages={115541},
  year={2024},
  publisher={Elsevier}
}

@article{shi2022analysis,
  title={Analysis of Q-learning like algorithms through evolutionary game dynamics},
  author={Shi, Yiming and Rong, Zhihai},
  journal={IEEE Transactions on Circuits and Systems II: Express Briefs},
  volume={69},
  number={5},
  pages={2463--2467},
  year={2022},
  publisher={IEEE}
}

@article{hu2021adaptive,
  title={Adaptive reputation promotes trust in social networks},
  author={Hu, Zhengyang and Li, Xiaopeng and Wang, Juan and Xia, Chengyi and Wang, Zhen and Perc, Matja{\v{z}}},
  journal={IEEE Transactions on Network Science and Engineering},
  volume={8},
  number={4},
  pages={3087--3098},
  year={2021},
  publisher={IEEE}
}

@article{li2022n,
  title={${N}$-Player Trust Game With Second-Order Reputation Evaluation in the Networked Population},
  author={Li, Xuezhu and Feng, Meiling and Han, Weiwei and Xia, Chengyi},
  journal={IEEE Systems Journal},
  volume={17},
  number={2},
  pages={2982--2992},
  year={2022},
  publisher={IEEE}
}

@article{zhu2025adaptive,
  title={Adaptive reputation promotes the cooperation of multiplayer snowdrift game on higher-order networks},
  author={Zhu, Yuying and Cui, Heng and Xia, Chengyi},
  journal={Physica A: Statistical Mechanics and its Applications},
  pages={130824},
  year={2025},
  publisher={Elsevier}
}

@article{zhu2025q,
  title={Q-learning update with second-order reputation promotes the evolution of trust within structured populations},
  author={Zhu, Yuying and Xing, Bohua and Xia, Chengyi},
  journal={Chaos, Solitons \& Fractals},
  volume={199},
  pages={116653},
  year={2025},
  publisher={Elsevier}
}

@article{helms2012keeping,
  title={Keeping up with the Joneses: Neighborhood effects in housing renovation},
  author={Helms, Andrew C},
  journal={Regional Science and Urban Economics},
  volume={42},
  number={1-2},
  pages={303--313},
  year={2012},
  publisher={Elsevier}
}

@article{tran2020my,
  title={My neighborhood has a good reputation: Associations between spatial stigma and health},
  author={Tran, Emma and Blankenship, Kim and Whittaker, Shannon and Rosenberg, Alana and Schlesinger, Penelope and Kershaw, Trace and Keene, Danya},
  journal={Health \& place},
  volume={64},
  pages={102392},
  year={2020},
  publisher={Elsevier}
}

@article{tanabe2013indirect,
  title={Indirect reciprocity with trinary reputations},
  author={Tanabe, Shoma and Suzuki, Hideyuki and Masuda, Naoki},
  journal={Journal of theoretical biology},
  volume={317},
  pages={338--347},
  year={2013},
  publisher={Elsevier}
}

@article{cuesta2015reputation,
  title={Reputation drives cooperative behaviour and network formation in human groups},
  author={Cuesta, Jose A and Gracia-L{\'a}zaro, Carlos and Ferrer, Alfredo and Moreno, Yamir and S{\'a}nchez, Angel},
  journal={Scientific reports},
  volume={5},
  number={1},
  pages={7843},
  year={2015},
  publisher={Nature Publishing Group UK London}
}

@article{zeng2025evolutionary,
  author={Zeng, Ziyan and Feng, Minyu and Szolnoki, Attila},
  journal={IEEE Transactions on Network Science and Engineering}, 
  title={Evolutionary Dynamics with Self-Interaction Learning in Networked Systems}, 
  year={2025},
  volume={},
  number={},
  pages={1-15},
  doi={10.1109/TNSE.2025.3583016}
}

@article{meng2024dynamics,
  title={Dynamics of collective cooperation under personalised strategy updates},
  author={Meng, Yao and Cornelius, Sean P and Liu, Yang-Yu and Li, Aming},
  journal={Nature Communications},
  volume={15},
  number={1},
  pages={3125},
  year={2024},
  publisher={Nature Publishing Group UK London}
}

@article{sheng2024strategy,
  title={Strategy evolution on higher-order networks},
  author={Sheng, Anzhi and Su, Qi and Wang, Long and Plotkin, Joshua B},
  journal={Nature Computational Science},
  volume={4},
  number={4},
  pages={274--284},
  year={2024},
  publisher={Nature Publishing Group US New York}
}

@article{meng2025promoting,
  title={Promoting collective cooperation through temporal interactions},
  author={Meng, Yao and McAvoy, Alex and Li, Aming},
  journal={Proceedings of the National Academy of Sciences},
  volume={122},
  number={26},
  pages={e2509575122},
  year={2025},
  publisher={National Academy of Sciences}
}

@article{li2025epidemic,
  title={Epidemic dynamics in homes and destinations under recurrent mobility patterns},
  author={Li, Yusheng and Yao, Yichao and Feng, Minyu and Benko, Tina P and Perc, Matja{\v{z}} and Zavr{\v{s}}nik, Jernej},
  journal={Chaos, Solitons \& Fractals},
  volume={195},
  pages={116273},
  year={2025},
  publisher={Elsevier}
}

@article{sheng2023evolutionary,
  title={Evolutionary dynamics on sequential temporal networks},
  author={Sheng, Anzhi and Li, Aming and Wang, Long},
  journal={PLOS Computational Biology},
  volume={19},
  number={8},
  pages={e1011333},
  year={2023},
  publisher={Public Library of Science San Francisco, CA USA}
}

@article{pi2025dynamic,
  title={Dynamic Evolution of Complex Networks: A Reinforcement Learning Approach Applying Evolutionary Games to Community Structure},
  author={Pi, Bin and Deng, Liang-Jian and Feng, Minyu and Perc, Matja{\v{z}} and Kurths, J{\"u}rgen},
  journal={IEEE Transactions on Pattern Analysis and Machine Intelligence}, 
  year={2025},
  volume={47},
  number={10},
  pages={8563-8582},
  publisher={IEEE},
  doi={10.1109/TPAMI.2025.3579895}
}

@article{li2020evolution,
  title={Evolution of cooperation on temporal networks},
  author={Li, Aming and Zhou, Lei and Su, Qi and Cornelius, Sean P and Liu, Yang-Yu and Wang, Long and Levin, Simon A},
  journal={Nature communications},
  volume={11},
  number={1},
  pages={2259},
  year={2020},
  publisher={Nature Publishing Group UK London}
}

%

\begin{IEEEbiography}[{\includegraphics[width=1in,height=1.25in,clip,keepaspectratio]{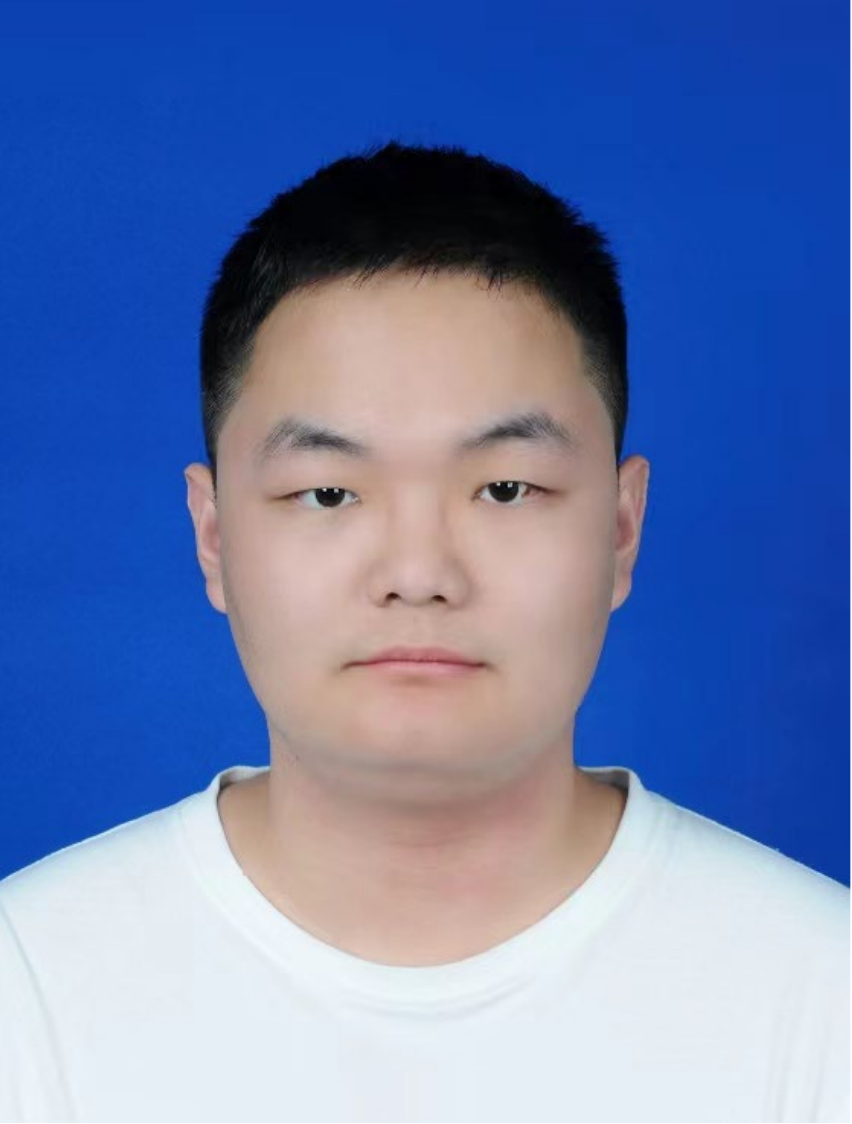}}]{Yuji Zhang}
received the B.E. degree from the College of Artificial Intelligence, Southwest University, Chongqing, China. He is currently pursuing an M.S. degree in electronic information. His research interests include evolutionary game theory, complex networks, network dynamics and their applications.
\end{IEEEbiography}

\begin{IEEEbiography}[{\includegraphics[width=1in,height=1.25in,clip,keepaspectratio]{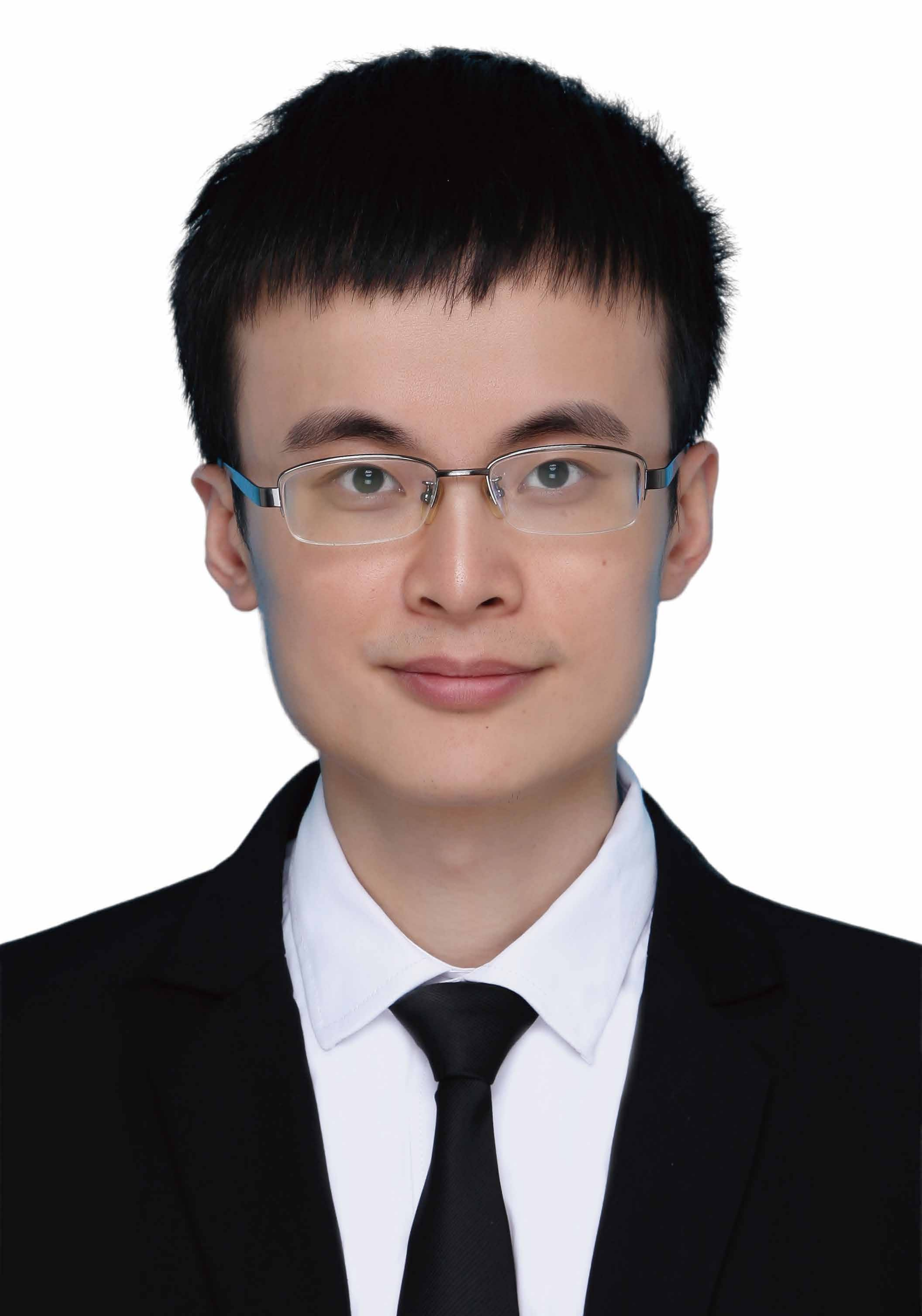}}]{Minyu Feng (IEEE Senior Member)}
received his Ph.D. degree in Computer Science from a joint program between University of Electronic Science and Technology of China, Chengdu, China, and Humboldt University of Berlin, Berlin, Germany, in 2018. Since 2019, he has been an associate professor at the College of Artificial Intelligence, Southwest University, Chongqing, China. 

Dr. Feng has published more than 70 peer-reviewed papers in authoritative journals, such as IEEE Transactions on Pattern Analysis and Machine Intelligence, IEEE Transactions on Systems, Man, and Cybernetics: Systems, IEEE Transactions on Cybernetics, etc. He is a Senior Member of China Computer Federation (CCF) and Chinese Association of Automation (CAA). Currently, he serves as a Subject Editor for Applied Mathematical Modelling, an Editorial Advisory Board Member for Chaos, and an Editorial Board Member for Humanities \& Social Sciences Communications, Scientific Reports, and International Journal of Mathematics for Industry. Besides, he is a Reviewer for Mathematical Reviews of the American Mathematical Society.

Dr. Feng's research interests include Complex Systems, Evolutionary Game Theory, Computational Social Science, and Mathematical Epidemiology.
\end{IEEEbiography}

\begin{IEEEbiography}[{\includegraphics[width=1in,height=1.25in,clip,keepaspectratio]{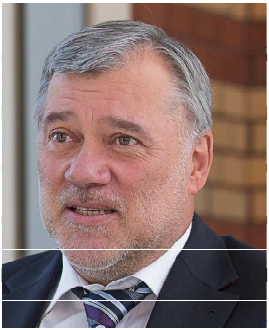}}]{J{\"u}rgen Kurths}
received the B.S. degree in mathematics from the University of Rostock, Rostock, Germany, the Ph.D. degree from the Academy of Sciences, German Democratic Republic, Berlin, Germany, in 1983, the Honorary degree from N.I.Lobachevsky State University, Nizhny Novgorod, Russia, in 2008, and the Honorary degree from Saratow State University, Saratov, Russia, in 2012. From 1994 to 2008, he was a Full Professor with the University of Potsdam, Potsdam, Germany. Since 2008, he has been a Professor of Nonlinear Dynamics with the Humboldt University of Berlin, Berlin, Germany, and the Chair of the Research Domain Complexity Science with the Potsdam Institute for Climate Impact Research, Potsdam. He has authored more than 700 papers, which are cited more than 60,000 times (H-index: 111). His main research interests include synchronization, complex networks, time series analysis, and their applications. Dr. Kurths was the recipient of the Alexander von Humboldt Research Award from India, in 2005, and from Poland in 2021, the Richardson Medal of the European Geophysical Union in 2013, and the Eight Honorary Doctorates. He is a Highly Cited Researcher in Engineering. He is a member of the Academia 1024 Europaea. He is an Editor-in-Chief of Chaos and on the editorial boards of more than ten journals. He is a Fellow of the American Physical Society, the Royal Society of 1023 Edinburgh, and the Network Science Society.
\end{IEEEbiography}

\begin{IEEEbiography}[{\includegraphics[width=1in,height=1.25in,clip,keepaspectratio]{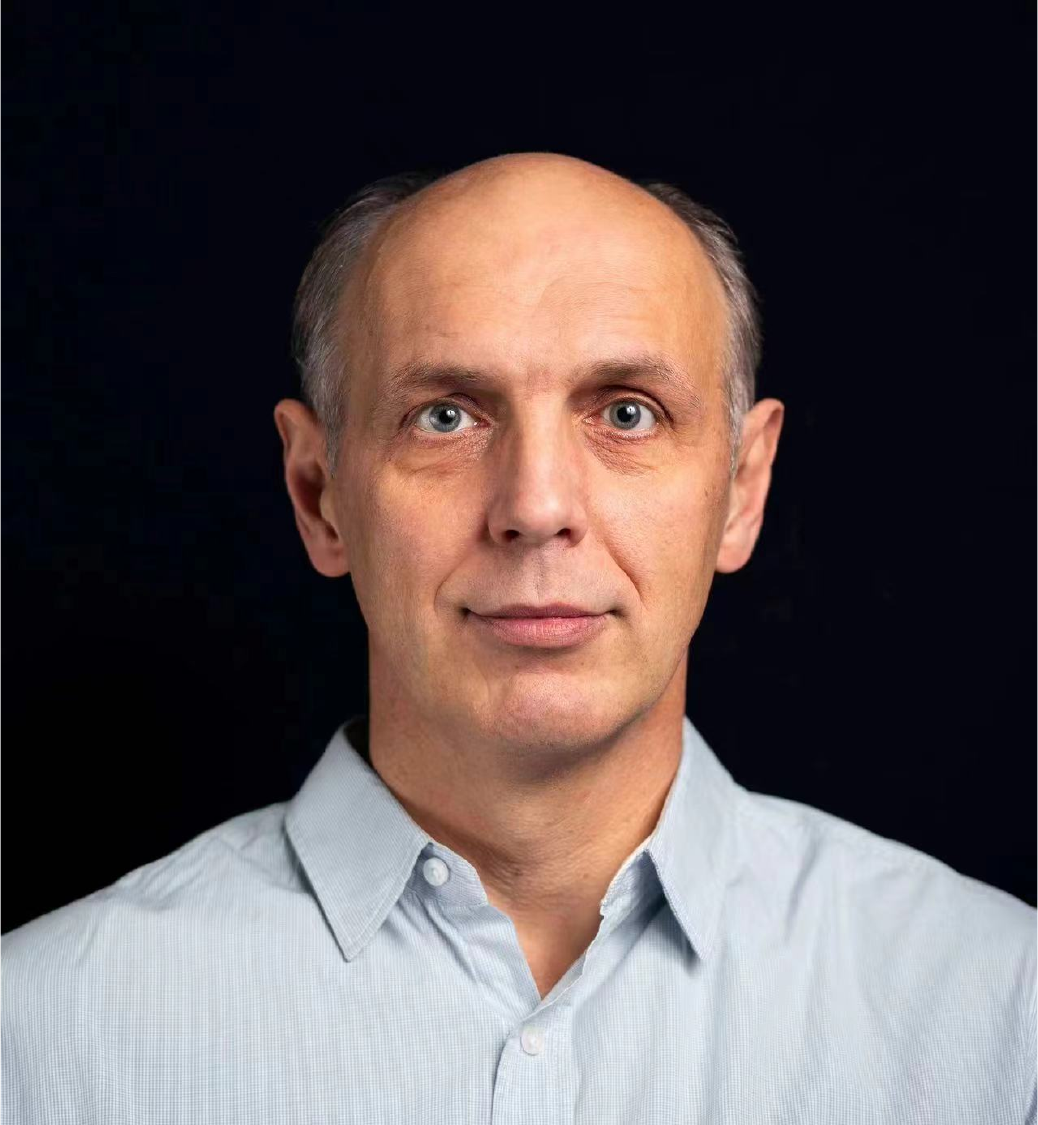}}]{Attila Szolnoki}
Attila Szolnoki is currently a research advisor with the Centre for Energy Research, Budapest, Hungary. His research interests include evolutionary game theory, phase transitions, and statistical physics and their applications. He is an Outstanding or a Distinguished Referee of several internationally recognized journals. He is/was an Editor of scientific journals including PNAS-Nexus, EPL, Physica A, Scientific Reports, Applied Mathematics and Computation, Frontiers in Physics, PLoS One, Entropy, and Indian Journal of Physics. He has authored around 200 original research papers with more than 25,000 citations and an H-factor of 80. He is among top 1\% most cited physicists according to Thomson Reuters Highly Cited Researchers.
\end{IEEEbiography}







\end{document}